\documentclass[aps, prl, 10pt, twocolumn, superscriptaddress, noshowpacs, preprintnumbers, longbibliography, bibnotes,hyperref,floatfix]{revtex4-1}

\usepackage[colorlinks=true,breaklinks=true]{hyperref}
\hypersetup{allcolors=[rgb]{0.0 0.0 1.0},linkcolor=[rgb]{0.75 0.05 0.05}}
\usepackage[dvipsnames]{xcolor}
\usepackage{mathtools}
\usepackage{amsmath}
\usepackage{amsfonts}
\usepackage{amssymb}
\usepackage{bbold}
\usepackage{multirow}
\usepackage{bm}
\usepackage{hyperref}
\long\def\exclude#1{}
\usepackage{orcidlink}
\usepackage{slashed}

\newcommand{\df}[1]{{\color{teal}{[{\bf DF:} #1]}}}

\newcommand{\cy}[1]{\textcolor{brown}{#1}}
\newcommand{\gcy}[1]{\textcolor{brown}{ (CY: #1) }}

\newcommand{\bp}{{\bf p}}

\newcommand{\br}{{\bf r}}
\newcommand{\bk}{{\bf k}}
\newcommand{\bq}{{\bf q}}

\begin{document}

\title{Presupernovae Strongly Constrain Electrophilic MeV Bosons}

\author{Yu Cheng
\orcidlink{0000-0002-4822-3890}}
\email{chengyu059@uchicago.edu}
\affiliation{Department of Physics, Enrico Fermi Institute, Leinweber Institute for Theoretical Physics,
Kavli Institute for Cosmological Physics, University of Chicago, Chicago, IL 60637, USA}
\affiliation{Department of Physics, Korea Advanced Institute of Science and Technology (KAIST), Daejeon 34141, Korea}

\author{Damiano F.\ G.\ Fiorillo \orcidlink{0000-0003-4927-9850}}
\email{damiano.fiorillo@gssi.it}
\affiliation{Gran Sasso Science Institute (GSSI), L’Aquila, Italy}

\author{Chui-Fan Kong \orcidlink{0009-0007-7010-5085}}
\email{kongcf@ibs.re.kr}
\affiliation{Particle Theory and Cosmology Group (PTC), Center for Theoretical Physics of the Universe (CTPU),
Institute for Basic Science, Daejeon 34126, Korea}

\author{Edoardo Vitagliano
\orcidlink{0000-0001-7847-1281}}
\email{edoardo.vitagliano@unipd.it}
\affiliation{Dipartimento di Fisica e Astronomia, Università degli Studi di Padova,
Via Marzolo 8, 35131 Padova, Italy}
\affiliation{Istituto Nazionale di Fisica Nucleare (INFN), Sezione di Padova,
Via Marzolo 8, 35131 Padova, Italy}

\preprint{CTPU-PTC-26-30}

\begin{abstract}
The years preceding a supernova (SN) can constrain new electrophilic bosons with MeV masses more than the explosion itself. Although cooler than SNe, the massive progenitors maintain consistently high temperatures much longer,  injecting more particles whose decays to positrons contribute to the diffuse Galactic 511-keV emission. For dark photons, the increase is exacerbated by the lower plasma frequency compared to SNe, where MeV excitations are screened. We find preSNe bounds on dark photon couplings up to two orders of magnitude stronger than SNe for masses between 1 and 3~MeV, with similar improvements for other electrophilic species such as axion-like particles.
\end{abstract}
\maketitle

\textbf{\textit{Introduction}}---Astrophysical environments act as natural factories for feebly interacting particles (FIPs) due to their extreme core temperatures and densities. If the plasma thermal kinetic energy reaches the kinematic threshold, meaning that the collision energy is high enough to supply the rest mass of the FIP, the novel particle can be produced. Ordinary stars with initial mass $\lesssim8M_\odot$, while being excellent laboratories for very light FIPs (see e.g.~\cite{Raffelt:1996wa}), never become efficient sources for particles heavier than several hundred keV, since they never reach temperatures $\gtrsim 10\,\rm keV$; instead, heavier stars can notoriously result in compact transients, namely supernovae (SNe) and neutron-star mergers, which are extremely powerful probes. In the last decade, a whole array of observables from these environments have been used to constrain e.g. axion-like particles (ALPs)~\cite{Giannotti:2010ty,Jaeckel:2017tud,Chang:2018rso,Calore:2020tjw,Croon:2020lrf,Calore:2021klc,Caputo:2021kcv,Caputo:2021rux,Mori:2021pcv,Caputo:2022rca,Lella:2022uwi,Hoof:2022xbe,Caputo:2022mah,Ferreira:2022xlw,Diamond:2023cto,Dev:2023hax,Fiorillo:2025yzf,Candon:2025ypl,Diamond:2023scc,Benabou:2024jlj,Fiorillo:2025sln,Chauhan:2025xqr,Bose:2026krk,Cui:2026wxv} (see also~\cite{Mikaelian:1978jg,Schramm:1981mk,Berezhiani:1999qh} for early attempts), millicharged particles~\cite{Davidson:2000hf,Chang:2018rso,Fiorillo:2024upk,Li:2024pcp}, majoron-like particles~\cite{Fiorillo:2022cdq,Fiorillo:2023ytr,Akita:2023iwq,Fiorillo:2023cas,Wang:2025qap,Telalovic:2024cot}, and dark photons (DPs)~\cite{Kazanas:2014mca,Rrapaj:2015wgs,Chang:2016ntp,Hardy:2016kme,DeRocco:2019njg,Sung:2019xie,Rrapaj:2019eam,Shin:2021bvz,Sieverding:2021jfa,Diamond:2021ekg,Caputo:2025avc,Caputo:2025aac,Mori:2025baz,Cheng:2026ugs,Cheng:2026zdc,Wang:2026zaq}.

\begin{figure}[!t]
\centering
\includegraphics[width=0.486\textwidth]{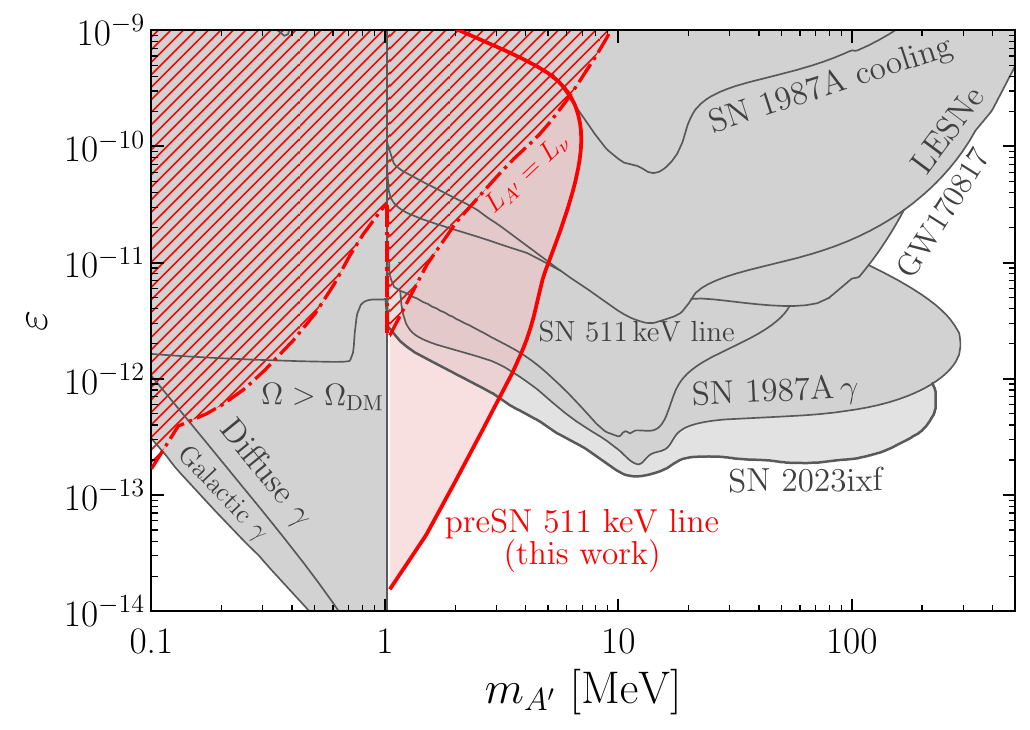}
\caption{The pink region shows the parameter space excluded by the Galactic 511~keV line, based on DP emission from massive stars before core collapse. 
The predicted signal is averaged over the IMF from $9 M_\odot$ to $60 M_\odot$.
We adopt the Ferri\`ere spiral-arm distribution with smearing length $\lambda_{\rm sm} = 1$~kpc. 
Above the dash-dotted line, $L_{A^\prime}$ exceeds neutrino luminosity at some stage of the stellar evolution, so DP emission cannot necessarily be treated as a perturbation (hatched). 
The grey regions show the limit reviewed in~\cite{Caputo:2025avc},
and the SN~2023ixf limit~\cite{Cheng:2026ugs}.
}
\label{fig:boundsIMF}
\end{figure}
%

%

A SN reaches its tens-of-MeV temperature through the collapse of the iron core after nuclear fusion has stopped (see e.g.~\cite{Burrows:1986me,Raffelt:1996wa}). Ostensibly, MeV FIPs are produced most efficiently \textit{after} collapse. However, years \textit{before} the collapse, the internal temperature had already reached hundreds of keV~\cite{Woosley:2002zz,Kippenhahn:2012qhp}, allowing for abundant FIP production~\cite{Rrapaj:2019eam, Sieverding:2021jfa}. Although significantly colder, this configuration is maintained for a much longer time, with a sustained emission that, for masses below several MeV, largely exceeds the subsequent SN one---and that could potentially impact the robustness of SN bounds themselves. The challenge thus becomes: How can we probe the copious emission of novel particles from preSNe? 

For electrophilic species, we propose using the particle decay as an observable. In this \textit{Letter}, we will use the paradigmatic example of DP, whose emissivity is further enhanced during heavy element burning compared to the SN phase, as large densities can hinder their production; in the supplemental material (SupM)~\cite{supplementalmaterial} we show that our results apply in a similar manner to ALP coupling to electrons. After being produced in preSNe, DPs undergo subsequent decays that result in an abundant injection of rather slow positrons. The latter lose their energy through Coulomb scattering with ambient electrons, emit bremsstrahlung radiation, and when nonrelativistic they form positronium bound states;  eventually, they annihilate and give rise to a 511 keV line~\cite{Prantzos:2010wi}. Compared to similar proposals in the context of SN-sourced FIPs~\cite{Calore:2021klc,Calore:2021lih}, the gargantuan boost in the putative flux can be used to explore uncharted regions of the parameter space. Using observations of the diffuse Galactic 511-keV emission from the INTErnational Gamma-Ray Astrophysics Laboratory (INTEGRAL) mission~\cite{Strong:2005zx,Bouchet:2008rp,Siegert:2015knp,Siegert:2019tus,Siegert:2022jii}, we find that the absence of such preSN-sourced flux can constrain MeV dark photons with $\varepsilon\simeq 10^{-14}$, surpassing existing bounds by as much as two orders of magnitude (see Fig.~\ref{fig:boundsIMF}).

\textbf{\textit{Production of dark photons in presupernovae}}---The dark photon is an additional Abelian gauge boson, denoted by $A'$, that interacts with the
standard model through kinetic mixing with the ordinary photon, $\mathcal{L}=(\varepsilon/2)F^{\mu\nu}F'_{\mu\nu}$, where $F'_{\mu\nu}=\partial_{\mu}A'_\nu-\partial_{\nu}A'_\mu$ and $\varepsilon$ is a dimensionless parameter~\cite{Okun:1982xi,Galison:1983pa,Holdom:1985ag,Holdom:1986eq} (see also~\cite{Dienes:1996zr,Goodsell:2009xc,Fabbrichesi:2020wbt,Albertus:2026fbe}). We work in natural units with $\hbar=c=1$. The drive of our work is the prominent emission of DPs with masses of a few MeV from stars nearing a SN explosion. We prove our point following the evolutionary track of massive stars from pre-main sequence to the onset of iron-core infall. We develop in parallel a parametric model of emission, informed by these simulations, that approximately reproduces the emission analytically.

We evolve a grid of stars with initial masses between $9$ and $60\,M_\odot$ at solar metallicity $Z = 0.02$ using MESA r26.4.1~\cite{Paxton:2010ji,Paxton:2013pj} until the core infall velocity is $100\,$km/s. 
From carbon burning onwards, we extract profiles of density $\rho$, temperature $T$, electron fraction $Y_e$.
Stars with initial masses up to $24 \, M_{\odot}$ end as red supergiants (RSGs) with photospheric radii $R_{\star} \simeq (600 - 1100) \, R_\odot$ in terms of the solar radius $R_\odot$.
Stars with initial masses above $25 \, M_{\odot}$ lose most of their hydrogen envelopes and end as compact helium supergiants with $R_{\star} \lesssim 3 R_\odot$.

A prototypical star, with initial mass $20\,M_\odot$, reaches core collapse as a $10.6\,M_\odot$ RSG. Central carbon burning occurs $10^2$--$10^3$ years before collapse at $T \simeq 80\,\mathrm{keV}$ and $\rho \simeq 10^5 \,\mathrm{g\,cm^{-3}}$. Oxygen burning proceeds during the final months at $T \simeq 190\,\mathrm{keV}$ and $\rho \simeq 4 \times 10^6 \,\mathrm{g\,cm^{-3}}$, followed by silicon burning during the final few days at $T \simeq 320\,\mathrm{keV}$ and $\rho \simeq 4 \times 10^7 \,\mathrm{g\,cm^{-3}}$ (see Fig.~\ref{fig:sm-maps} in the  SupM). 
Thus, electron density reaches $n_e\sim \rho/m_p\sim 10^{30}$--$10^{31}\,\mathrm{cm}^{-3}$ in terms of proton mass $m_p$, 
with a plasma frequency $\omega_{\rm pl} = \sqrt{4\pi \alpha n_e/m_e}\sim 30$--$70\,\mathrm{keV}$.
In the SN cores, by contrast,
$\omega_{\rm pl}\sim 10\,\mathrm{MeV}$, suppressing emission of DPs with lower masses which are rapidly screened by the plasma. In practice, the effective in-medium mixing $\varepsilon_{\mathrm{eff},T}\simeq \varepsilon m_{A'}^2/\lvert m_{A'}^2-\omega_{\mathrm{pl}}^2\rvert$ is suppressed in SNe by a factor $m_{A'}^2/\omega_{\mathrm{pl}}^2$ for $m_{A'}\sim 1\,\mathrm{MeV}$ and $\omega_{\mathrm{pl}}\sim 10\,\mathrm{MeV}$. No such suppression occurs in preSNe.

\begin{figure}[!t]
\centering
\includegraphics[width=0.486\textwidth]{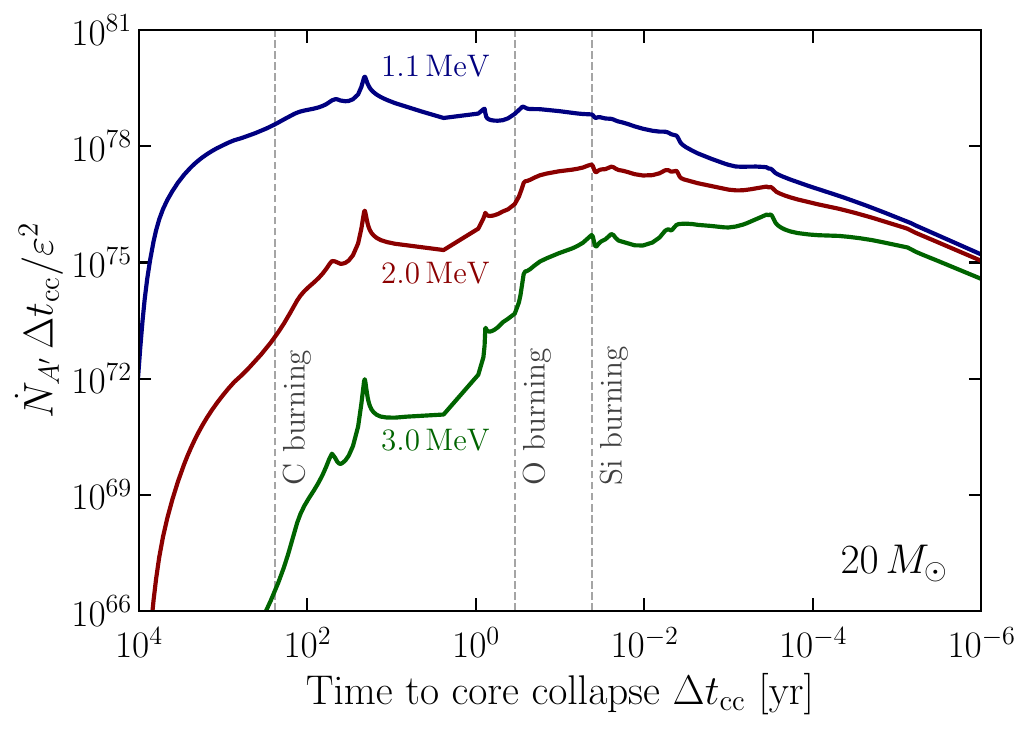}
\caption{DP production rate multiplied by the time to core collapse, $\Delta t_{\rm cc}$, normalized by $\varepsilon^{2}$, as a function of $\Delta t_{\rm cc}$ for a star with mass $M_\star = 20 M_\odot$.
The dashed lines mark the onset of 
the central carbon-, oxygen-, and silicon-burning stages, respectively. 
}
\label{fig:ratettM20}
\end{figure}

In these hot stellar cores DPs are emitted primarily by coalescence $e^+ e^-\to A'$. The emission rate is elementary and reviewed in our SupM: we can assume leptons to have Maxwell-Boltzmann distributions, since only the high-energy tail of their distribution can emit MeV-scale DPs. Then the number of DPs emitted is
\begin{equation}\label{eq:emissivity_exact}
    \frac{dN_{A'}}{dVdt}=\frac{\alpha \varepsilon^2 (m_{A'}^2+2m_e^2)T \sqrt{m_{A'}^2-4m_e^2}}{2\pi^2}K_1\left(\frac{m_{A'}}{T}\right),
\end{equation}
where $\alpha$ is the fine-structure constant and
$K_1(x)$ is the modified Bessel function of the second kind of order one. 
Numerically, we use the full Fermi–Dirac expression in Eq.~\eqref{eq:sm-pair-rate}, integrated over the stellar volume, throughout the stellar evolution to determine the cumulative DP emission. 
In Fig.~\ref{fig:ratettM20}, we show $\dot N_{A'}\Delta t_{\rm cc}$ as a function of the time to core collapse, $\Delta t_{\rm cc}$, estimating the number of DPs emitted over the relevant timescale.
For DP masses up to a few MeV, the emitted number peaks at a few years to few days before collapse, as anticipated. 
In fact, the quantitative emission can be understood rather easily using the asymptotic expansion of the  $K_1(x)$ for $x\gg 1$.
Over the final year, during which the central temperature of the benchmark star rises from $110$ to $320$\,keV, we can model the emitting core as a single zone of mass $M_{\rm c}\simeq 1.5 M_\odot$ with constant $T \simeq 160$~keV and $\rho \simeq 6 \times10^6\,\mathrm{g/cm^3}$.
Within a time $\Delta t_{\rm cc}\sim 1\,\mathrm{yr}$, we find a total number of emitted DPs $\mathcal{N}\sim M_{\rm c}\Delta t_{\rm cc}\rho^{-1} dN_{A'}/dVdt$ of order
\begin{align}
    \frac{\mathcal{N}}{\varepsilon^2} 
    \sim 1.7\times 10^{82}  
    &\left(\frac{m_{A'}}{m_e}\right)^{\! 5/2}\sqrt{1-\frac{4m_e^2}{m_{A'}^2}} \nonumber
    \\
   &\times \left(1+\frac{2m_e^2}{m_{A'}^2}\right)\exp\left(-\frac{3.2 \, m_{A'}}{m_e}\right),
    \label{eq:TotalnumberDP}
\end{align}
in remarkable agreement with Fig.~\ref{fig:ratettM20}.
Note here that the DP luminosity $L_{A'}$ that accompanies this emission is negligible compared to the neutrino luminosity $L_\nu$ of the star for the couplings of interest. 
For $\varepsilon = 10^{-12}$ and $m_{A'} = 2\,$MeV, 
the luminosity ratio $L_{A'}/L_{\nu}$
of the benchmark star remains below approximately $10^{-2}$ throughout the entire course of its late evolution (see SupM Fig.~\ref{fig:sm-maps}), 
so the standard stellar evolution is unaffected.

\textit{\textbf{Positrons from dark photon decays}}---
Each DP eventually decays into an electron–positron pair at the rate
\begin{equation}
    \Gamma_{A'}=\frac{\alpha \varepsilon^2 m_{A'}}{3}\left(1+\frac{2m_e^2}{m_{A'}^2}\right)\sqrt{1-\frac{4m_e^2}{m_{A'}^2}}.
\label{eq:GammaAp}
\end{equation}
the DP with momentum $k$ then travels a typical decay length $\ell_{\rm dec} = k/(m_{A'} \Gamma_{A'})$.
For the thermally distributed DPs, we can define the mean free path as $\langle \ell_{\rm dec} \rangle = \langle v \rangle/\Gamma_{A'}$, where $\langle v\rangle=\sqrt{8T/\pi m_{A'}}$ is the average velocity at the temperature $T$ of the emitting plasma.

For this estimate, we use the one-zone model introduced above with $T\simeq160$~keV, giving $\langle v\rangle \simeq 0.45$ for $m_{A'}=2$~MeV. 
For $\varepsilon = 10^{-13}$ and $m_{A'} = 2\,$MeV, 
the typical mean free path is $\langle \ell_{\rm dec} \rangle \simeq 2 \times 10^{17}\,{\rm cm}\,\simeq 0.06\,$pc,
well beyond the largest stellar radius in our models, $R_\star \simeq 8 \times 10^{13} \,\mathrm{cm}$, 
but still well within the Galaxy.
Only at larger mixings $\varepsilon\gtrsim10^{-11}$, a sizable fraction of DPs decays within the progenitor. Their positrons then remain trapped and do not contribute to the observable signal.

\exclude{
(\cy{Old one}DPs decay into electrons and positrons with a mean free path $\ell_{A'}=(\Gamma_{A'} \langle v\rangle)^{-1}$, where $\langle v\rangle=\sqrt{8T/\pi m_{A'}}$ is the average velocity for the thermally distributed DPs---for this approximate calculation it is enough to consider the simple one-zone model. The decay rate
\begin{equation}
    \Gamma_{A'}=\frac{\alpha \varepsilon^2 m_{A'}}{3}\left(1+\frac{2m_e^2}{m_{A'}^2}\right)\sqrt{1-\frac{4m_e^2}{m_{A'}^2}}.
\end{equation}
DPs decaying within the star radius $R^*=XXX$ do not lead to a visible signal since the positrons are trapped inside of it )
\df{Should we comment on other possible observables as e.g. Chauhan et al. or circumstellar medium?}. 
}

Positrons are born with kinetic energies between
a few tens of keV and about 1 MeV. 
In the interstellar medium (ISM) with a typical density $n\simeq 1\,\mathrm{cm}^{-3}$, the mildly relativistic positrons primarily lose energy through ionization. 
The stopping power can be extracted e.g. from Ref.~\cite{NIST_ESTAR} as $S_e \simeq 4\,\mathrm{MeV\, cm}^2/\mathrm{g}$. 
Therefore, the typical time for a positron moving with the speed of light to lose its energy $E\simeq 1\,\mathrm{MeV}$ is
\begin{equation}
    \tau\simeq \frac{E}{S_e n_H m_p }\simeq 10^{13}\,\mathrm{s}.
\end{equation}
where $m_p$ is the proton mass.
Once a positron has stopped, 
it can annihilate on a timescale shorter than or comparable to its stopping time, 
primarily through positronium formation with electrons in the ISM~\cite{Prantzos:2010wi},
resulting in the emission of 511-keV line.
Para-positronium accounts for a fraction $k_{\rm ps}=1/4$ of the positronium formed and decays into two $511$~keV photons, yielding an average of $2k_{\rm ps}=1/2$ line photons per positron. 
For positrons with initial energies in the MeV range, annihilation in flight before stopping removes only $0.2$--$3\%$ of the positrons, a small correction that we neglect.

Within a time $\tau$, positrons spread spatially. Their gyroradius in the Galactic magnetic field is so small that gyroresonant diffusion is not obvious, since turbulent fluctuations at small scales are likely damped. 
Positrons instead are stuck to the field line they gyrate around; these field lines induce a random motion whose diffusion coefficient dimensionally is of the form $D\sim (\delta B/B)^2\ell_c/3$ where $\ell_c$ is the coherence length and $\delta B/B$ is the fractional turbulent fluctuation. With $\ell_c\sim 10-100\,\mathrm{pc}$ and $\delta B/B\sim 0.1$, 
we can estimate $D\sim 10^{28-29}\,\mathrm{cm^2/s}$, so that the spatial diffusion of positrons roughly spans a few hundred parsecs.
If we convolve the Gaussian diffusion kernel $e^{-r^2/4 D t}$ with the exponential distribution of the positron lifetimes,  $ e^{-t / \tau }/\tau$, 
we find that the distribution of the annihilation point of positrons is smeared through a normalized distribution
\begin{equation}
    P(\br)
    =
    \frac{1}{4\pi |\br|\lambda^2_{\rm sm}} 
    e^{-|\br|/\lambda_{\rm sm}},
    \label{eq:smearingL}
\end{equation}
with $\lambda_{\rm sm} \equiv \sqrt{D \tau}$ the smearing length.
The typical smearing distance is of the order of 100~pc, and is thus small compared with Galactic scales.
In practice, we follow Refs.~\cite{Calore:2021lih,DelaTorreLuque:2023huu} and apply the smearing separately to the source distribution in the Galactic plane and perpendicular to it, 
with exponential kernels of scale $\lambda_{\rm sm}$ in each direction, reaching up to a conservative value of 1~kpc:
for details, see the SupM.

\textbf{\textit{The 511-keV line and Constraint}}---
Since the smearing in Eq.~\eqref{eq:smearingL} already accounts for the positron propagation, 
what remains to be computed for each progenitor is  the number of positrons it injects into the ISM that eventually annihilate at rest, locally on Galactic scales.
For a progenitor with initial mass $M$, we calculate the number of positrons that eventually annihilate locally by integrating its DP emission over the pre-SN evolution,
\begin{equation}
    N_{\rm loc}=\int\! dt\,\dot N_{A'}(t)\left\langle e^{-R_\star(t)/\ell_{\rm dec}(k)}
    \bar{P}_{\rm surv}(k)\right\rangle_{k},
    \label{eq:NlocProd}
\end{equation}
where $\langle\cdots\rangle_k$ averages over the DP spectrum emitted at time $t$, $e^{-R_\star/\ell_{\rm dec}}$ is the fraction of DPs decaying outside the star and $\bar{P}_{\rm surv} \simeq0 .97$--$1$ the probability that the positron is not annihilated in flight~\cite{Beacom:2005qv}.
The latter depends on the positron energy, which for an isotropic decay is uniformly distributed between  $(\omega \mp k \beta) / 2$, 
with $\omega$ the DP energy and $\beta=\sqrt{1-4 m_e^2 / m_{A^{\prime}}^2}$ (see also Eq.~\eqref{eq:sm-psurv-avg} in SupM).
We then average these yields over the progenitor initial mass function (IMF), which gives the number of stars born per unit initial mass. Adopting the Salpeter form $dN/dM\propto M^{-2.35}$~\cite{Salpeter:1955it,Baldry:2003xi} over the mass range $9$--$60\,M_\odot$, we obtain the mean positron yield per progenitor,
\begin{equation}
    \langle N_{\rm loc}\rangle
    =
    \frac{\int_{9\,M_{\odot}}^{60\,M_{\odot}} dM\,M^{-2.35}\,N_{\rm loc} (M)}{\int_{9\,M_{\odot}}^{60\,M_{\odot}} dM\,M^{-2.35}},
\end{equation}
where $N_{\rm loc}(M)$ interpolates between the stellar models.

The positron annihilation timescale, $\sim10^5$~yr, is much longer than the interval between Galactic core collapses, so the Galaxy hosts a steady positron population. 
The resulting $511$~keV intensity toward Galactic longitude $l$ and latitude $b$ is
\begin{equation}
    \frac{d\phi_{511}}{d\Omega}(l,b)
    =
    \frac{2k_{\rm ps}}{4\pi}\,
    \langle N_{\rm loc} \rangle\,
    \Gamma_{\rm cc}\int ds\, \tilde{n}_{cc}\!\left[\br(s,l,b)\right],
    \label{eq:flux}
\end{equation}
where $s$ is the distance along the line of sight, 
$N_{\rm loc}$ the number of positrons per collapse that annihilate locally, 
$\Gamma_{\rm cc}$ the Galactic core-collapse rate, 
and $\tilde{n}_{cc}$ the spatial distribution of annihilation sites, normalized to unity.
Since the sources are stars approaching core collapse, we adopt the corresponding SN rate, $\Gamma_{\rm cc}=2.3$ per century~\cite{Li:2010kd}, 
and the three-dimensional SN distribution implemented in DRAGON2~\cite{Evoli:2016xgn}. 
This model combines the Ferriere distribution of Ref.~\cite{Ferriere:2001rg,Siegert:2015knp} with the four-arm spiral pattern of Ref.~\cite{Steiman-Cameron:2010iuq}. Following Refs.~\cite{Calore:2021lih,DelaTorreLuque:2023huu}, we obtain $\tilde{n}_{cc}$ by smearing its source distribution $n_{cc}$ over the benchmark propagation length $\lambda_{\rm sm}=1$~kpc (see details in the SupM).

We compare the prediction with the SPI longitude profile of the line of Ref.~\cite{Siegert:2015knp}, measured in $3^\circ$ bins over $-10.75^\circ< b <10.25^\circ$. 
The pre-SN sources trace the Galactic disk, 
so their predicted longitude profile is nearly flat out to $|l| \simeq 50^\circ$. 
The observed emission, however, peaks in the bulge and is about an order of magnitude fainter in the disk.
As the pre-SN contribution increases, 
it therefore first exceeds the data in the faint disk bins at $|l| \gtrsim 10^\circ$.
To avoid background-subtraction systematics in the outer bins, 
we restrict the analysis to $|l|<20^\circ$ and adopt the one-sided $2\sigma$ criterion of Ref.~\cite{DelaTorreLuque:2023huu},
\begin{equation}
    \chi^2 
    \equiv
    \sum_{i,\ |l_i|<20^\circ}\left(\frac{\max\left[\Phi_i^{\rm pred}-\Phi_i,\,0\right]}{\sigma_i}\right)^2\le 4,
    \label{eq:criterion}
\end{equation}
where $\Phi_i^{\rm pred}$ is Eq.~\eqref{eq:flux} integrated over the $i$th bin and $\Phi_i$ is the measured flux with standard deviation $\sigma_i$. 
Since this criterion counts only predicted excesses, without modelling the bulge or subtracting known disk sources, 
it gives a conservative bound on the positron yield, 
independently of the particle model. 
The number of positrons per progenitor that annihilate locally is limited to $N_{\rm max}=2.0\times10^{52}$ 
for $\lambda_{\rm sm} = 1\,$kpc.

Figure~\ref{fig:boundsIMF} shows the resulting $2\sigma$ exclusion region through the $\chi^2$ fit, 
which is equivalent to $\langle N_{\rm loc} \rangle > N_{\rm max}$. With this condition, we find, using $\mathcal{N}$ from Eq.~\eqref{eq:TotalnumberDP}, that 
\begin{align}
    \varepsilon \lesssim 10^{-15}\,\exp &\left(1.6\, \frac{m_{A'}}{m_e}\right) \left(\frac{m_e}{m_{A'}}\right)^{5/4}
    \nonumber
    \\
    &\times{\left(1-\frac{4m_e^2}{m_{A'}^2}\right)^{\!\!-1/4}\left(1+\frac{2m_e^2}{m_{A'}^2}\right)^{\!\!-1/2} },
\end{align}
in remarkable agreement with our numerical constraint at $m_{A'}\simeq 1-2\,\mathrm{MeV}$. At higher masses, the numerical bounds are stronger, since they benefit from the emission at times closer to the SN explosion which are not captured by our one-zone model.

Our new bounds supersede the ones obtained from SN~2023ixf and SN~1987A by one to two orders of magnitude for $1\,\mathrm{MeV}\lesssim m_{A'}\lesssim 3\,\mathrm{MeV}$. For the first time, we thus find that the preSN can constrain new species beyond the reach of the SN itself. The upper edge of our constrained region descends from the decay length becoming shorter than the stellar radius. In addition, we assume that DP cooling is subdominant to the
neutrino cooling: this condition holds
below the dash-dotted line $L_{A'}=L_\nu$, where DP emission can be treated as a perturbation to standard stellar evolution~\cite{Rrapaj:2019eam,Sieverding:2021jfa}.
Note that both $L_{A'}$ and $L_\nu$ evolve with time, and we take the 
maximum value of $L_{A'}/L_\nu$ during the whole evolution process to be 1 as the perturbation condition.
While the lower edge lies deep within this perturbative regime ($L_{A'}/L_\nu \le 2 \times 10^{-4}$), 
DP emission above the $L_{A'}=L_\nu$ line may alter the stellar evolution, making the unperturbed stellar models unreliable. This curve then is a conservative upper boundary of our exclusion, together with the trapping ceiling.

The lower boundary is robust against the stellar model and Galactic SN distributions.
DP production is governed by the core temperature–density track during carbon,
oxygen, and silicon burning and changes little between the different MESA setups,
despite their different envelopes.
Different Galactic SN spatial distributions also give nearly identical bounds: a detailed comparison is in Fig.~\ref{fig:sm-rate-compare} and Fig.~\ref{fig:sm-bounds-models} of the SupM.


The existence of dense circumstellar material (CSM) can  reduce the diffuse signal. 
More than $30$\% of Type~II SN progenitors are surrounded by CSM extending to $r_{\rm CSM}\simeq10^{14}$--$10^{15}$~cm because of increased mass loss shortly before explosion~\cite{Forster:2018mib,Bruch:2020jcr,Jacobson-Galan:2021pki,Bruch:2022aqd}. 
Positrons from DPs decaying inside the CSM stop and annihilate there, 
producing a transient $511$~keV signal before the SN~\cite{Cheng:2026zdc} (see also Refs.~\cite{Cheng:2026ugs,Chauhan:2025xqr}), instead of contributing to the diffuse line. For these stars, $r_{\rm CSM}$ replaces $R_\star$ in Eq.~\eqref{eq:NlocProd}.
At the lower boundary, 
$\ell_{\rm dec}\simeq10$~pc, $0.05$~pc, and $4\times10^{15}$~cm for $m_{A'}=1.1$, $2$, and $3$~MeV, respectively. For $r_{\rm CSM}=10^{15}$~cm, about $7.5$\% of the RSG positrons are lost at $3$~MeV, weakening the bound by $3$\%.
The change stays below $1$\% for $m_{A'}\le2.5$~MeV and for the smaller CSM radius $r_{\rm CSM}=10^{14}$~cm.

\textbf{\textit{Discussion and outlook}}---As massive stars approach the end of their lives, their cores contract and heat up to 
sustain advanced nuclear fusion stages, resulting in 
temperatures and densities orders of magnitude higher 
than those found in younger stars. Although cooler than 
SNe, such heavy stars are exceptionally efficient 
factories for the production of hypothetical FIPs, such as 
DPs, thanks to their years-long emission. In this \textit{Letter}, we have shown 
that the large DP emission leads to a positron flux ensuing an unavoidable 
contribution to the 511 keV line. Using the diffuse 
Galactic 511-keV emission measurements 
by the spectrometer on INTEGRAL (SPI), we found new 
constraints on MeV DPs as powerful as $\varepsilon\lesssim 10^{-14}$. 

In our SupM we show that 
similarly competitive bounds can be obtained on ALPs coupling to electrons, 
reaching $g_{ae}\lesssim 10^{-14}$. These findings can 
be generalized to other leptophilic FIPs, e.g. sterile neutrinos, which feature a 
decay channel $\nu_s\rightarrow \nu_a e^+ e^-$ ($\nu_a$ being an active neutrino).  Furthermore, while we focused on electrophilic bosons, other heavy FIPs also are copiously produced in the final stages of their evolution. In that case, the observable effect would not be the positron annihilation, but might come from other effects, e.g.\ the decay into electromagnetic radiation, a possibility that we plan to explore in future work.

Thus, while previous 
observables leveraged helium burning to probe particles as 
heavy as several hundred keV~\cite{Candon:2024eah,Fiorillo:2026shb,Fiorillo:2026vkd,Fiorillo:2026psi}, we studied heavy-element ($\rm C,O,Si$) burning stages and established stringent bounds on electrophilic bosons with masses up to  several MeV. Our results can be contrasted with Refs.~\cite{Sieverding:2021jfa}, that considered preSN neutrinos from a star at a distance of $500\,\rm pc$ and found projections reaching $\epsilon\lesssim 10^{-12}$; and Refs.~\cite{Heger:2008er,Dominguez:2025bgg}, in which the 
impact of massless particles was considered, and ultimately found negligible when 
accounting for existing constraints.

Finally, the large emission of novel species suggests a nonperturbative 
impact on the star well before it reaches the SN stage, even 
in parameter regions where their impact on the SN has 
traditionally been treated as perturbative. Although we 
cannot establish how existing SN bounds are affected, our 
findings reveal a potentially important systematic 
uncertainty that must be considered when assessing their robustness.


\textbf{\textit{Acknowledgments---}} 
Y.C. is supported by the National Research Foundation of Korea (NRF) Grant RS-2023-00211732, 
by the Samsung Science and Technology Foundation under Project Number SSTF-BA2302-05, by the POSCO Science Fellowship of POSCO TJ Park Foundation, and by the NRF Grant RS-2024-00405629. D.F.G.F. and E.V. acknowledge support from Istituto Nazionale di Fisica Nucleare (INFN) through the Theoretical Astroparticle Physics (TASP) project.
C.-F.K. is supported by IBS under the project code, IBS-R018-D1. 
E.V. acknowledges support from the Italian Ministero dell'Università e della Ricerca through the FIS 2 project FIS-2023-01577 (DD n. 23314 10-12-2024, CUP C53C24001460001) and through Departments of Excellence grant 2023--2027 ``Quantum Frontier''. 
Funded by the European Union (ERC, NEPTUNE, No. 101301124). Views and opinions expressed are however those of the author(s) only and do not necessarily reflect those of the European Union or the European Research Council. Neither the European Union nor the granting authority can be held responsible for them. The ERC funding acknowledged here supports the contribution of D.F.G.F. to this work.

\bibliographystyle{bibi}
\bibliography{References}

@misc{supplementalmaterial,
  note = {See the Supplemental Material for additional discussions on the DP production from preSN core, MESA simulations and stellar model comparisons, spatial and mass distributions of preSNe, positron survival probability, meared 511 keV photon flux and SPI dataset, a comparison of DP constraints from different preSNe profiles and positron smearing scales, and ALP production and constraints. }
}

@article{Paxton:2013pj,
    author = "Paxton, Bill and others",
    title = "{Modules for Experiments in Stellar Astrophysics (MESA): Planets, Oscillations, Rotation, and Massive Stars}",
    eprint = "1301.0319",
    archivePrefix = "arXiv",
    primaryClass = "astro-ph.SR",
    doi = "10.1088/0067-0049/208/1/4",
    journal = "Astrophys. J. Suppl.",
    volume = "208",
    pages = "4",
    year = "2013"
}

@article{Paxton:2010ji,
    author = "Paxton, Bill and Bildsten, Lars and Dotter, Aaron and Herwig, Falk and Lesaffre, Pierre and Timmes, Frank",
    collaboration = "MESA",
    title = "{Modules for Experiments in Stellar Astrophysics (MESA)}",
    eprint = "1009.1622",
    archivePrefix = "arXiv",
    primaryClass = "astro-ph.SR",
    doi = "10.1088/0067-0049/192/1/3",
    journal = "Astrophys. J. Suppl.",
    volume = "192",
    pages = "3",
    year = "2011"
}

@article{Bruch:2022aqd,
    author = "Bruch, Rachel J. and others",
    title = "{The Prevalence and Influence of Circumstellar Material around Hydrogen-rich Supernova Progenitors}",
    eprint = "2212.03313",
    archivePrefix = "arXiv",
    primaryClass = "astro-ph.HE",
    doi = "10.3847/1538-4357/acd8be",
    journal = "Astrophys. J.",
    volume = "952",
    number = "2",
    pages = "119",
    year = "2023"
}

@article{Jacobson-Galan:2021pki,
    author = "Jacobson-Gal{\'a}n, Wynn and others",
    title = "{Final Moments. I. Precursor Emission, Envelope Inflation, and Enhanced Mass Loss Preceding the Luminous Type II Supernova 2020tlf}",
    eprint = "2109.12136",
    archivePrefix = "arXiv",
    primaryClass = "astro-ph.HE",
    doi = "10.3847/1538-4357/ac3f3a",
    journal = "Astrophys. J.",
    volume = "924",
    number = "1",
    pages = "15",
    year = "2022"
}

@article{Bruch:2020jcr,
    author = "Bruch, Rachel J. and others",
    title = "{A Large Fraction of Hydrogen-rich Supernova Progenitors Experience Elevated Mass Loss Shortly Prior to Explosion}",
    eprint = "2008.09986",
    archivePrefix = "arXiv",
    primaryClass = "astro-ph.HE",
    doi = "10.3847/1538-4357/abef05",
    journal = "Astrophys. J.",
    volume = "912",
    number = "1",
    pages = "46",
    year = "2021"
}

@article{Forster:2018mib,
    author = {F{\"o}rster, F. and others},
    title = "{The delay of shock breakout due to circumstellar material evident in most type II supernovae}",
    eprint = "1809.06379",
    archivePrefix = "arXiv",
    primaryClass = "astro-ph.HE",
    doi = "10.1038/s41550-018-0563-4",
    journal = "Nature Astron.",
    volume = "2",
    number = "10",
    pages = "808--818",
    year = "2018",
    note = "[Erratum: Nature Astron. 3, 107 (2019)]"
}

@article{Sieverding:2021jfa,
    author = "Sieverding, A. and Rrapaj, E. and Guo, G. and Qian, Y. -Z.",
    title = "{Impact of Dark Photon Emission on Massive Star Evolution and Pre-Supernova Neutrino Signal}",
    eprint = "2101.08672",
    archivePrefix = "arXiv",
    primaryClass = "astro-ph.SR",
    doi = "10.3847/1538-4357/abee84",
    journal = "Astrophys. J.",
    volume = "912",
    number = "1",
    pages = "13",
    year = "2021"
}

@article{Siegert:2015knp,
    author = "Siegert, Thomas and Diehl, Roland and Khachatryan, Gerasim and Krause, Martin G. H. and Guglielmetti, Fabrizia and Greiner, Jochen and Strong, Andrew W. and Zhang, Xiaoling",
    title = "{Gamma-ray spectroscopy of Positron Annihilation in the Milky Way}",
    eprint = "1512.00325",
    archivePrefix = "arXiv",
    primaryClass = "astro-ph.HE",
    doi = "10.1051/0004-6361/201527510",
    journal = "Astron. Astrophys.",
    volume = "586",
    pages = "A84",
    year = "2016"
}

@article{DelaTorreLuque:2023huu,
    author = "De la Torre Luque, Pedro and Balaji, Shyam and Carenza, Pierluca",
    title = "{Multimessenger search for electrophilic feebly interacting particles from supernovae}",
    eprint = "2307.13731",
    archivePrefix = "arXiv",
    primaryClass = "hep-ph",
    doi = "10.1103/PhysRevD.109.103028",
    journal = "Phys. Rev. D",
    volume = "109",
    number = "10",
    pages = "103028",
    year = "2024"
}

@article{Steiman-Cameron:2010iuq,
    author = "Steiman-Cameron, Thomas Y. and Wolfire, Mark and Hollenbach, David",
    title = "{COBE and the Galactic Interstellar Medium: Geometry of the Spiral Arms from FIR Cooling Lines}",
    doi = "10.1088/0004-637X/722/2/1460",
    journal = "Astrophys. J.",
    volume = "722",
    pages = "1460--1473",
    year = "2010"
}

@article{Ferriere:2001rg,
    author = "Ferriere, Katia M.",
    title = "{The interstellar environment of our galaxy}",
    eprint = "astro-ph/0106359",
    archivePrefix = "arXiv",
    doi = "10.1103/RevModPhys.73.1031",
    journal = "Rev. Mod. Phys.",
    volume = "73",
    pages = "1031--1066",
    year = "2001"
}

@article{Evoli:2016xgn,
    author = "Evoli, Carmelo and Gaggero, Daniele and Vittino, Andrea and Di Bernardo, Giuseppe and Di Mauro, Mattia and Ligorini, Arianna and Ullio, Piero and Grasso, Dario",
    title = "{Cosmic-ray propagation with $\small{DRAGON2}$: I. numerical solver and astrophysical ingredients}",
    eprint = "1607.07886",
    archivePrefix = "arXiv",
    primaryClass = "astro-ph.HE",
    reportNumber = "TUM-HEP-1056-16",
    doi = "10.1088/1475-7516/2017/02/015",
    journal = "JCAP",
    volume = "02",
    pages = "015",
    year = "2017"
}

@article{Li:2010kd,
    author = "Li, Weidong and Chornock, Ryan and Leaman, Jesse and Filippenko, Alexei V. and Poznanski, Dovi and Wang, Xiaofeng and Ganeshalingam, Mohan and Mannucci, Filippo",
    title = "{Nearby Supernova Rates from the Lick Observatory Supernova Search. III. The Rate-Size Relation, and the Rates as a Function of Galaxy Hubble Type and Colour}",
    eprint = "1006.4613",
    archivePrefix = "arXiv",
    primaryClass = "astro-ph.SR",
    doi = "10.1111/j.1365-2966.2011.18162.x",
    journal = "Mon. Not. Roy. Astron. Soc.",
    volume = "412",
    pages = "1473",
    year = "2011"
}

@article{Baldry:2003xi,
    author = "Baldry, Ivan K. and Glazebrook, Karl",
    title = "{Constraints on a universal IMF from UV to near-IR galaxy luminosity densities}",
    eprint = "astro-ph/0304423",
    archivePrefix = "arXiv",
    doi = "10.1086/376502",
    journal = "Astrophys. J.",
    volume = "593",
    pages = "258--271",
    year = "2003"
}

@article{Beacom:2005qv,
    author = "Beacom, John F. and Yuksel, Hasan",
    title = "{Stringent constraint on galactic positron production}",
    eprint = "astro-ph/0512411",
    archivePrefix = "arXiv",
    doi = "10.1103/PhysRevLett.97.071102",
    journal = "Phys. Rev. Lett.",
    volume = "97",
    pages = "071102",
    year = "2006"
}

@article{Caputo:2025avc,
    author = "Caputo, Andrea and Park, Jaeyoung and Yun, Seokhoon",
    title = "{Heavy dark photon handbook: Cosmological and astrophysical bounds}",
    eprint = "2511.15785",
    archivePrefix = "arXiv",
    primaryClass = "hep-ph",
    reportNumber = "CTPU-PTC-25-42",
    doi = "10.1103/m6zs-jxxp",
    journal = "Phys. Rev. D",
    volume = "113",
    number = "7",
    pages = "075001",
    year = "2026"
}

@article{Cheng:2026ugs,
    author = "Cheng, Yu and Kong, Chui-Fan and Lin, Yen-Hsun and Wu, Meng-Ru and Yun, Seokhoon",
    title = "{Circumstellar Medium of Supernovae as New Probes for Feebly Interacting Particles}",
    eprint = "2603.09615",
    archivePrefix = "arXiv",
    primaryClass = "hep-ph",
    doi = "10.1103/8v59-c52b",
    journal = "Phys. Rev. Lett.",
    volume = "137",
    number = "11",
    pages = "111005",
    year = "2026"
}

@article{Calore:2020tjw,
    author = "Calore, Francesca and Carenza, Pierluca and Giannotti, Maurizio and Jaeckel, Joerg and Mirizzi, Alessandro",
    title = "{Bounds on axionlike particles from the diffuse supernova flux}",
    eprint = "2008.11741",
    archivePrefix = "arXiv",
    primaryClass = "hep-ph",
    doi = "10.1103/PhysRevD.102.123005",
    journal = "Phys. Rev. D",
    volume = "102",
    number = "12",
    pages = "123005",
    year = "2020"
}

@article{Candon:2024eah,
    author = "Cand{\'o}n, Francisco R. and Fiorillo, Damiano F. G. and Lucente, Giuseppe and Vitagliano, Edoardo and Vogel, Julia K.",
    title = "{NuSTAR Bounds on Radiatively Decaying Particles from M82}",
    eprint = "2412.03660",
    archivePrefix = "arXiv",
    primaryClass = "hep-ph",
    doi = "10.1103/PhysRevLett.134.171004",
    journal = "Phys. Rev. Lett.",
    volume = "134",
    number = "17",
    pages = "171004",
    year = "2025"
}

@article{Jaeckel:2017tud,
    author = "Jaeckel, J. and Malta, P. C. and Redondo, J.",
    title = "{Decay photons from the axionlike particles burst of type II supernovae}",
    eprint = "1702.02964",
    archivePrefix = "arXiv",
    primaryClass = "hep-ph",
    doi = "10.1103/PhysRevD.98.055032",
    journal = "Phys. Rev. D",
    volume = "98",
    number = "5",
    pages = "055032",
    year = "2018"
}

@book{Raffelt:1996wa,
    author = "Raffelt, G. G.",
    title = "{Stars as laboratories for fundamental physics}: {The astrophysics of neutrinos, axions, and other weakly interacting particles}",
    isbn = "978-0-226-70272-8",
    month = "5",
    year = "1996"
}

@article{Calore:2021lih,
    author = "Calore, Francesca and Carenza, Pierluca and Giannotti, Maurizio and Jaeckel, Joerg and Lucente, Giuseppe and Mastrototaro, Leonardo and Mirizzi, Alessandro",
    title = "{511~keV line constraints on feebly interacting particles from supernovae}",
    eprint = "2112.08382",
    archivePrefix = "arXiv",
    primaryClass = "hep-ph",
    doi = "10.1103/PhysRevD.105.063026",
    journal = "Phys. Rev. D",
    volume = "105",
    number = "6",
    pages = "063026",
    year = "2022"
}

@article{Sung:2019xie,
    author = "Sung, Allan and Tu, Huitzu and Wu, Meng-Ru",
    title = "{New constraint from supernova explosions on light particles beyond the Standard Model}",
    eprint = "1903.07923",
    archivePrefix = "arXiv",
    primaryClass = "hep-ph",
    doi = "10.1103/PhysRevD.99.121305",
    journal = "Phys. Rev. D",
    volume = "99",
    number = "12",
    pages = "121305",
    year = "2019"
}

@article{Dev:2023hax,
    author = "Dev, P. S. Bhupal and Fortin, Jean-Fran\c{c}ois and Harris, Steven P. and Sinha, Kuver and Zhang, Yongchao",
    title = "{First Constraints on the Photon Coupling of Axionlike Particles from Multimessenger Studies of the Neutron Star Merger GW170817}",
    eprint = "2305.01002",
    archivePrefix = "arXiv",
    primaryClass = "hep-ph",
    reportNumber = "INT-PUB-23-014",
    doi = "10.1103/PhysRevLett.132.101003",
    journal = "Phys. Rev. Lett.",
    volume = "132",
    number = "10",
    pages = "101003",
    year = "2024"
}

@article{Chang:2018rso,
    author = "Chang, Jae Hyeok and Essig, Rouven and McDermott, Samuel D.",
    title = "{Supernova 1987A Constraints on Sub-GeV Dark Sectors, Millicharged Particles, the QCD Axion, and an Axion-like Particle}",
    eprint = "1803.00993",
    archivePrefix = "arXiv",
    primaryClass = "hep-ph",
    reportNumber = "YITP-SB-18-01, FERMILAB-PUB-17-432-T",
    doi = "10.1007/JHEP09(2018)051",
    journal = "JHEP",
    volume = "09",
    pages = "051",
    year = "2018"
}

@article{Croon:2020lrf,
    author = "Croon, Djuna and Elor, Gilly and Leane, Rebecca K. and McDermott, Samuel D.",
    title = "{Supernova Muons: New Constraints on $Z$' Bosons, Axions and ALPs}",
    eprint = "2006.13942",
    archivePrefix = "arXiv",
    primaryClass = "hep-ph",
    reportNumber = "MIT-CTP/5214, FERMILAB-PUB-20-246-A-T",
    doi = "10.1007/JHEP01(2021)107",
    journal = "JHEP",
    volume = "01",
    pages = "107",
    year = "2021"
}

@article{Ferreira:2022xlw,
    author = {Ferreira, Ricardo Z. and Marsh, M. C. David and M\"uller, Eike},
    title = "{Strong supernovae bounds on ALPs from quantum loops}",
    eprint = "2205.07896",
    archivePrefix = "arXiv",
    primaryClass = "hep-ph",
    doi = "10.1088/1475-7516/2022/11/057",
    journal = "JCAP",
    volume = "11",
    pages = "057",
    year = "2022"
}

@article{Hoof:2022xbe,
    author = "Hoof, Sebastian and Schulz, Lena",
    title = "{Updated constraints on axion-like particles from temporal information in supernova SN1987A gamma-ray data}",
    eprint = "2212.09764",
    archivePrefix = "arXiv",
    primaryClass = "hep-ph",
    reportNumber = "TTP22-072",
    doi = "10.1088/1475-7516/2023/03/054",
    journal = "JCAP",
    volume = "03",
    pages = "054",
    year = "2023"
}

@article{Fiorillo:2024upk,
    author = "Fiorillo, Damiano F. G. and Vitagliano, Edoardo",
    title = "{Self-Interacting Dark Sectors in Supernovae Can Behave as a Relativistic Fluid}",
    eprint = "2404.07714",
    archivePrefix = "arXiv",
    primaryClass = "hep-ph",
    doi = "10.1103/PhysRevLett.133.251004",
    journal = "Phys. Rev. Lett.",
    volume = "133",
    number = "25",
    pages = "251004",
    year = "2024"
}

@article{Diamond:2023scc,
    author = "Diamond, Melissa and Fiorillo, Damiano F. G. and Marques-Tavares, Gustavo and Vitagliano, Edoardo",
    title = "{Axion-sourced fireballs from supernovae}",
    eprint = "2303.11395",
    archivePrefix = "arXiv",
    primaryClass = "hep-ph",
    doi = "10.1103/PhysRevD.107.103029",
    journal = "Phys. Rev. D",
    volume = "107",
    number = "10",
    pages = "103029",
    year = "2023",
    note = "[Erratum: Phys.Rev.D 108, 049902 (2023)]"
}

@article{Fiorillo:2022cdq,
    author = "Fiorillo, Damiano F. G. and Raffelt, Georg G. and Vitagliano, Edoardo",
    title = "{Strong Supernova 1987A Constraints on Bosons Decaying to Neutrinos}",
    eprint = "2209.11773",
    archivePrefix = "arXiv",
    primaryClass = "hep-ph",
    doi = "10.1103/PhysRevLett.131.021001",
    journal = "Phys. Rev. Lett.",
    volume = "131",
    number = "2",
    pages = "021001",
    year = "2023"
}

@article{Fiorillo:2023ytr,
    author = "Fiorillo, Damiano F. G. and Raffelt, Georg G. and Vitagliano, Edoardo",
    title = "{Large Neutrino Secret Interactions Have a Small Impact on Supernovae}",
    eprint = "2307.15115",
    archivePrefix = "arXiv",
    primaryClass = "hep-ph",
    doi = "10.1103/PhysRevLett.132.021002",
    journal = "Phys. Rev. Lett.",
    volume = "132",
    number = "2",
    pages = "021002",
    year = "2024"
}

@article{Fiorillo:2023cas,
    author = "Fiorillo, Damiano F. G. and Raffelt, Georg G. and Vitagliano, Edoardo",
    title = "{Supernova emission of secretly interacting neutrino fluid: Theoretical foundations}",
    eprint = "2307.15122",
    archivePrefix = "arXiv",
    primaryClass = "hep-ph",
    doi = "10.1103/PhysRevD.109.023017",
    journal = "Phys. Rev. D",
    volume = "109",
    number = "2",
    pages = "023017",
    year = "2024"
}

@article{Akita:2023iwq,
    author = "Akita, Kensuke and Im, Sang Hui and Masud, Mehedi and Yun, Seokhoon",
    title = "{Limits on heavy neutral leptons, Z' bosons and majorons from high-energy supernova neutrinos}",
    eprint = "2312.13627",
    archivePrefix = "arXiv",
    primaryClass = "hep-ph",
    reportNumber = "CTPU-PTC-23-55",
    doi = "10.1007/JHEP07(2024)057",
    journal = "JHEP",
    volume = "07",
    pages = "057",
    year = "2024"
}

@article{Giannotti:2010ty,
    author = "Giannotti, M. and Duffy, L. D. and Nita, R.",
    title = "{New constraints for heavy axion-like particles from supernovae}",
    eprint = "1009.5714",
    archivePrefix = "arXiv",
    primaryClass = "astro-ph.HE",
    reportNumber = "LA-UR-10-05895",
    doi = "10.1088/1475-7516/2011/01/015",
    journal = "JCAP",
    volume = "01",
    pages = "015",
    year = "2011"
}

@article{Benabou:2024jlj,
    author = "Benabou, Joshua N. and Manzari, Claudio Andrea and Park, Yujin and Prabhakar, Garima and Safdi, Benjamin R. and Savoray, Inbar",
    title = "{Time-delayed gamma-ray signatures of heavy axions from core-collapse supernovae}",
    eprint = "2412.13247",
    archivePrefix = "arXiv",
    primaryClass = "hep-ph",
    month = "12",
    year = "2024"
}

@article{Davidson:2000hf,
    author = "Davidson, Sacha and Hannestad, Steen and Raffelt, Georg",
    title = "{Updated bounds on millicharged particles}",
    eprint = "hep-ph/0001179",
    archivePrefix = "arXiv",
    reportNumber = "CERN-TH-99-384",
    doi = "10.1088/1126-6708/2000/05/003",
    journal = "JHEP",
    volume = "05",
    pages = "003",
    year = "2000"
}

@article{Lella:2022uwi,
    author = "Lella, Alessandro and Carenza, Pierluca and Lucente, Giuseppe and Giannotti, Maurizio and Mirizzi, Alessandro",
    title = "{Protoneutron stars as cosmic factories for massive axionlike particles}",
    eprint = "2211.13760",
    archivePrefix = "arXiv",
    primaryClass = "hep-ph",
    doi = "10.1103/PhysRevD.107.103017",
    journal = "Phys. Rev. D",
    volume = "107",
    number = "10",
    pages = "103017",
    year = "2023"
}

@article{Telalovic:2024cot,
    author = "Telalovic, Bernanda and Fiorillo, Damiano F. G. and Mart\'\i{}nez-Mirav\'e, Pablo and Vitagliano, Edoardo and Bustamante, Mauricio",
    title = "{The next galactic supernova can uncover mass and couplings of particles decaying to neutrinos}",
    eprint = "2406.15506",
    archivePrefix = "arXiv",
    primaryClass = "hep-ph",
    doi = "10.1088/1475-7516/2024/11/011",
    journal = "JCAP",
    volume = "11",
    pages = "011",
    year = "2024"
}

@article{Caputo:2021rux,
    author = "Caputo, Andrea and Raffelt, Georg and Vitagliano, Edoardo",
    title = "{Muonic boson limits: Supernova redux}",
    eprint = "2109.03244",
    archivePrefix = "arXiv",
    primaryClass = "hep-ph",
    reportNumber = "MPP-2021-154",
    doi = "10.1103/PhysRevD.105.035022",
    journal = "Phys. Rev. D",
    volume = "105",
    number = "3",
    pages = "035022",
    year = "2022"
}

@article{Diamond:2023cto,
    author = "Diamond, Melissa and Fiorillo, Damiano F. G. and Marques-Tavares, Gustavo and Tamborra, Irene and Vitagliano, Edoardo",
    title = "{Multimessenger Constraints on Radiatively Decaying Axions from GW170817}",
    eprint = "2305.10327",
    archivePrefix = "arXiv",
    primaryClass = "hep-ph",
    doi = "10.1103/PhysRevLett.132.101004",
    journal = "Phys. Rev. Lett.",
    volume = "132",
    number = "10",
    pages = "101004",
    year = "2024"
}

@article{Caputo:2022rca,
    author = "Caputo, Andrea and Raffelt, Georg and Vitagliano, Edoardo",
    title = "{Radiative transfer in stars by feebly interacting bosons}",
    eprint = "2204.11862",
    archivePrefix = "arXiv",
    primaryClass = "astro-ph.SR",
    doi = "10.1088/1475-7516/2022/08/045",
    journal = "JCAP",
    volume = "08",
    number = "08",
    pages = "045",
    year = "2022"
}

@article{Caputo:2022mah,
    author = "Caputo, Andrea and Janka, Hans-Thomas and Raffelt, Georg and Vitagliano, Edoardo",
    title = "{Low-Energy Supernovae Severely Constrain Radiative Particle Decays}",
    eprint = "2201.09890",
    archivePrefix = "arXiv",
    primaryClass = "astro-ph.HE",
    doi = "10.1103/PhysRevLett.128.221103",
    journal = "Phys. Rev. Lett.",
    volume = "128",
    number = "22",
    pages = "221103",
    year = "2022"
}

@article{Diamond:2021ekg,
    author = "Diamond, Melissa D. and Marques-Tavares, Gustavo",
    title = "{\ensuremath{\gamma}-Ray Flashes from Dark Photons in Neutron Star Mergers}",
    eprint = "2106.03879",
    archivePrefix = "arXiv",
    primaryClass = "hep-ph",
    doi = "10.1103/PhysRevLett.128.211101",
    journal = "Phys. Rev. Lett.",
    volume = "128",
    number = "21",
    pages = "211101",
    year = "2022"
}

@misc{NIST_ESTAR,
  author = {Berger, M. J. and Coursey, J. S. and
            Zucker, M. A. and Chang, J.},
  title = {{ESTAR, PSTAR, and ASTAR}: Computer
           Programs for Calculating Stopping-Power
           and Range Tables for Electrons, Protons,
           and Helium Ions},
  year = {2005},
  note = {Version 1.2.3, National Institute of
          Standards and Technology,
          Gaithersburg, MD},
  url = {https://physics.nist.gov/PhysRefData/Star/Text/ESTAR.html}
}

@article{Caputo:2021kcv,
    author = "Caputo, Andrea and Carenza, Pierluca and Lucente, Giuseppe and Vitagliano, Edoardo and Giannotti, Maurizio and Kotake, Kei and Kuroda, Takami and Mirizzi, Alessandro",
    title = "{Axionlike Particles from Hypernovae}",
    eprint = "2104.05727",
    archivePrefix = "arXiv",
    primaryClass = "hep-ph",
    doi = "10.1103/PhysRevLett.127.181102",
    journal = "Phys. Rev. Lett.",
    volume = "127",
    number = "18",
    pages = "181102",
    year = "2021"
}

@article{Braaten:1993jw,
    author = "Braaten, Eric and Segel, Daniel",
    title = "{Neutrino energy loss from the plasma process at all temperatures and densities}",
    eprint = "hep-ph/9302213",
    archivePrefix = "arXiv",
    primaryClass = "hep-ph",
    doi = "10.1103/PhysRevD.48.1478",
    journal = "Phys. Rev. D",
    volume = "48",
    pages = "1478--1491",
    year = "1993"
}

@article{Lorimer:2006qs,
    author = "Lorimer, D. R. and Faulkner, A. J. and Lyne, A. G. and Manchester, R. N. and Kramer, M. and McLaughlin, M. A. and Hobbs, G. and Possenti, A. and Stairs, I. H. and Camilo, F. and Burgay, M. and D'Amico, N. and Corongiu, A. and Crawford, F.",
    title = "{The Parkes multibeam pulsar survey: VI. Discovery and timing of 142 pulsars and a Galactic population analysis}",
    eprint = "astro-ph/0607640",
    archivePrefix = "arXiv",
    primaryClass = "astro-ph",
    doi = "10.1111/j.1365-2966.2006.10887.x",
    journal = "Mon. Not. Roy. Astron. Soc.",
    volume = "372",
    pages = "777--800",
    year = "2006"
}

@article{Salpeter:1955it,
    author = "Salpeter, Edwin E.",
    title = "{The Luminosity function and stellar evolution}",
    doi = "10.1086/145971",
    journal = "Astrophys. J.",
    volume = "121",
    pages = "161--167",
    year = "1955"
}

@article{Fiorillo:2025sln,
    author = "Fiorillo, Damiano F. G. and Pitik, Tetyana and Vitagliano, Edoardo",
    title = "{Supernova production of axionlike particles coupling to electrons, reloaded}",
    eprint = "2503.15630",
    archivePrefix = "arXiv",
    primaryClass = "hep-ph",
    doi = "10.1103/y1r2-gtb5",
    journal = "Phys. Rev. D",
    volume = "112",
    number = "8",
    pages = "083008",
    year = "2025"
}

@article{Wang:2025qap,
    author = "Wang, Isaac R. and Xu, Xun-Jie and Zhou, Bei",
    title = "{Widen the Resonance: Probing a New Regime of Neutrino Self-Interactions with Astrophysical Neutrinos}",
    eprint = "2501.07624",
    archivePrefix = "arXiv",
    primaryClass = "hep-ph",
    reportNumber = "FERMILAB-PUB-25-0016-T",
    doi = "10.1103/9ddp-j1z9",
    journal = "Phys. Rev. Lett.",
    volume = "135",
    number = "18",
    pages = "181002",
    year = "2025"
}

@article{Fiorillo:2025yzf,
    author = "Fiorillo, Damiano F. G. and Pitik, Tetyana and Vitagliano, Edoardo",
    title = "{Energy Transfer by Feebly Interacting Particles in Supernovae: The Trapping Regime}",
    eprint = "2503.13653",
    archivePrefix = "arXiv",
    primaryClass = "hep-ph",
    doi = "10.1103/cz94-dqxt",
    journal = "Phys. Rev. Lett.",
    volume = "135",
    number = "7",
    pages = "071005",
    year = "2025"
}

@article{Candon:2025ypl,
    author = "Cand{\'o}n, Francisco R. and Fiorillo, Damiano F. G. and Janka, Hans-Thomas and van Baal, Bart F. A. and Vitagliano, Edoardo",
    title = "{Small Progenitors, Large Couplings: Type Ic Supernova Constraints on Radiatively Decaying Particles}",
    eprint = "2509.18253",
    archivePrefix = "arXiv",
    primaryClass = "hep-ph",
    doi = "10.1103/d562-j1yz",
    journal = "Phys. Rev. Lett.",
    volume = "137",
    number = "6",
    pages = "061004",
    year = "2026"
}

@article{An:2013yfc,
    author = "An, Haipeng and Pospelov, Maxim and Pradler, Josef",
    title = "{New stellar constraints on dark photons}",
    eprint = "1302.3884",
    archivePrefix = "arXiv",
    primaryClass = "hep-ph",
    doi = "10.1016/j.physletb.2013.07.008",
    journal = "Phys. Lett. B",
    volume = "725",
    pages = "190--195",
    year = "2013"
}

@article{Redondo:2013lna,
    author = "Redondo, Javier and Raffelt, Georg",
    title = "{Solar constraints on hidden photons re-visited}",
    eprint = "1305.2920",
    archivePrefix = "arXiv",
    primaryClass = "hep-ph",
    doi = "10.1088/1475-7516/2013/08/034",
    journal = "JCAP",
    volume = "08",
    pages = "034",
    year = "2013"
}

@article{Chang:2016ntp,
    author = "Chang, Jae Hyeok and Essig, Rouven and McDermott, Samuel D.",
    title = "{Revisiting Supernova 1987A Constraints on Dark Photons}",
    eprint = "1611.03864",
    archivePrefix = "arXiv",
    primaryClass = "hep-ph",
    reportNumber = "YITP-SB-16-44",
    doi = "10.1007/JHEP01(2017)107",
    journal = "JHEP",
    volume = "01",
    pages = "107",
    year = "2017"
}

@article{Hardy:2016kme,
    author = "Hardy, Edward and Lasenby, Robert",
    title = "{Stellar cooling bounds on new light particles: plasma mixing effects}",
    eprint = "1611.05852",
    archivePrefix = "arXiv",
    primaryClass = "hep-ph",
    doi = "10.1007/JHEP02(2017)033",
    journal = "JHEP",
    volume = "02",
    pages = "033",
    year = "2017"
}

@article{Kazanas:2014mca,
    author = "Kazanas, Demos and Mohapatra, Rabindra N. and Nussinov, Shmuel and Teplitz, Vigdor L. and Zhang, Yongchao",
    title = "{Supernova Bounds on the Dark Photon Using its Electromagnetic Decay}",
    eprint = "1410.0221",
    archivePrefix = "arXiv",
    primaryClass = "hep-ph",
    reportNumber = "UMD-PP--014-015",
    doi = "10.1016/j.nuclphysb.2014.11.009",
    journal = "Nucl. Phys. B",
    volume = "890",
    pages = "17--29",
    year = "2014"
}

@article{DeRocco:2019njg,
    author = "DeRocco, William and Graham, Peter W. and Kasen, Daniel and Marques-Tavares, Gustavo and Rajendran, Surjeet",
    title = "{Observable signatures of dark photons from supernovae}",
    eprint = "1901.08596",
    archivePrefix = "arXiv",
    primaryClass = "hep-ph",
    doi = "10.1007/JHEP02(2019)171",
    journal = "JHEP",
    volume = "02",
    pages = "171",
    year = "2019"
}

@article{Cheng:2026zdc,
    author = "Cheng, Yu and Kong, Chui-Fan and Lin, Yen-Hsun and Wu, Meng-Ru and Yun, Seokhoon",
    title = "{Probing Feebly Interacting Particles with 511 keV Line from Circumstellar Medium of Supernovae}",
    eprint = "2609.18247",
    archivePrefix = "arXiv",
    primaryClass = "hep-ph",
    reportNumber = "CTPU-PTC-26-26",
    month = "9",
    year = "2026"
}

@article{Caputo:2025aac,
    author = "Caputo, Andrea and Janka, Hans-Thomas and Raffelt, Georg and Yun, Seokhoon",
    title = "{Cooling the Shock: New Supernova Constraints on Dark Photons}",
    eprint = "2502.01731",
    archivePrefix = "arXiv",
    primaryClass = "hep-ph",
    reportNumber = "CERN-TH-2025-020, CTPU-PTC-25-05",
    doi = "10.1103/PhysRevLett.134.151002",
    journal = "Phys. Rev. Lett.",
    volume = "134",
    number = "15",
    pages = "151002",
    year = "2025"
}

@article{Rrapaj:2015wgs,
    author = "Rrapaj, Ermal and Reddy, Sanjay",
    title = "{Nucleon-nucleon bremsstrahlung of dark gauge bosons and revised supernova constraints}",
    eprint = "1511.09136",
    archivePrefix = "arXiv",
    primaryClass = "nucl-th",
    reportNumber = "INT-PUB-15-065",
    doi = "10.1103/PhysRevC.94.045805",
    journal = "Phys. Rev. C",
    volume = "94",
    number = "4",
    pages = "045805",
    year = "2016"
}

@article{Shin:2021bvz,
    author = "Shin, Chang Sub and Yun, Seokhoon",
    title = "{Dark gauge boson production from neutron stars via nucleon-nucleon bremsstrahlung}",
    eprint = "2110.03362",
    archivePrefix = "arXiv",
    primaryClass = "hep-ph",
    doi = "10.1007/JHEP02(2022)133",
    journal = "JHEP",
    volume = "02",
    pages = "133",
    year = "2022"
}

@article{Chauhan:2025xqr,
    author = "Chauhan, Garv and Lunardini, Cecilia",
    title = "{511 keV Gamma Ray Echo from Particle Decays in Supernovae}",
    eprint = "2511.14857",
    archivePrefix = "arXiv",
    primaryClass = "hep-ph",
    month = "11",
    year = "2025"
}

@article{Bose:2026krk,
    author = "Bose, Debajit and Ganguly, Sougata and Kumar, Tanmoy and Raj, Nirmal",
    title = "{$\gamma$hosts from the past: reheating neutron stars with supernova-sourced ALPs}",
    eprint = "2610.30255",
    archivePrefix = "arXiv",
    primaryClass = "hep-ph",
    reportNumber = "CTPU-PTC-26-27",
    month = "10",
    year = "2026"
}

@article{Woosley:2002zz,
    author = "Woosley, S. E. and Heger, A. and Weaver, T. A.",
    title = "{The evolution and explosion of massive stars}",
    doi = "10.1103/RevModPhys.74.1015",
    journal = "Rev. Mod. Phys.",
    volume = "74",
    pages = "1015--1071",
    year = "2002"
}

@book{Kippenhahn:2012qhp,
    author = "Kippenhahn, Rudolf and Weigert, Alfred and Weiss, Achim",
    title = "{Stellar structure and evolution}",
    doi = "10.1007/978-3-642-30304-3",
    isbn = "978-0-387-58013-5, 978-3-642-30255-8, 978-3-642-30304-3",
    publisher = "Springer",
    series = "Astronomy and Astrophysics Library",
    month = "8",
    year = "2012"
}

@article{Burrows:1986me,
    author = "Burrows, Adam and Lattimer, James M.",
    title = "{The birth of neutron stars}",
    doi = "10.1086/164405",
    journal = "Astrophys. J.",
    volume = "307",
    pages = "178--196",
    year = "1986"
}

@article{Calore:2021klc,
    author = "Calore, Francesca and Carenza, Pierluca and Giannotti, Maurizio and Jaeckel, Joerg and Lucente, Giuseppe and Mirizzi, Alessandro",
    title = "{Supernova bounds on axionlike particles coupled with nucleons and electrons}",
    eprint = "2107.02186",
    archivePrefix = "arXiv",
    primaryClass = "hep-ph",
    doi = "10.1103/PhysRevD.104.043016",
    journal = "Phys. Rev. D",
    volume = "104",
    number = "4",
    pages = "043016",
    year = "2021"
}

@article{Strong:2005zx,
    author = "Strong, Andrew W. and Diehl, R. and Halloin, H. and Schoenfelder, V. and Bouchet, L. and Mandrou, P. and Lebrun, F. and Terrier, R.",
    title = "{Gamma-ray continuum emission from the inner galactic region as observed with integral/spi}",
    eprint = "astro-ph/0509290",
    archivePrefix = "arXiv",
    doi = "10.1051/0004-6361:20053798",
    journal = "Astron. Astrophys.",
    volume = "444",
    pages = "495",
    year = "2005"
}

@article{Bouchet:2008rp,
    author = "Bouchet, L. and Jourdain, E. and Roques, J. P. and Strong, A. and Diehl, R. and Lebrun, F. and Terrier, R.",
    title = "{INTEGRAL SPI All-Sky View in Soft Gamma Rays: Study of Point Source and Galactic Diffuse Emissions}",
    eprint = "0801.2086",
    archivePrefix = "arXiv",
    primaryClass = "astro-ph",
    doi = "10.1086/529489",
    journal = "Astrophys. J.",
    volume = "679",
    pages = "1315",
    year = "2008"
}

@article{Dirac:1930bga,
    author = "Dirac, P. A. M.",
    title = "{On the Annihilation of Electrons and Protons}",
    doi = "10.1017/S0305004100016091",
    journal = "Proc. Cambridge Phil. Soc.",
    volume = "26",
    pages = "361--375",
    year = "1930"
}

@article{Siegert:2019tus,
    author = "Siegert, Thomas and Crocker, Roland M. and Diehl, Roland and Krause, Martin G. H. and Panther, Fiona H. and Pleintinger, Moritz M. M. and Weinberger, Christoph",
    title = "{Constraints on positron annihilation kinematics in the inner Galaxy}",
    eprint = "1906.00498",
    archivePrefix = "arXiv",
    primaryClass = "astro-ph.HE",
    doi = "10.1051/0004-6361/201833856",
    journal = "Astron. Astrophys.",
    volume = "627",
    pages = "A126",
    year = "2019"
}

@article{Siegert:2022jii,
    author = "Siegert, Thomas and Berteaud, Joanna and Calore, Francesca and Serpico, Pasquale D. and Weinberger, Christoph",
    title = "{Diffuse Galactic emission spectrum between 0.5 and 8.0 MeV}",
    eprint = "2202.04574",
    archivePrefix = "arXiv",
    primaryClass = "astro-ph.HE",
    doi = "10.1051/0004-6361/202142639",
    journal = "Astron. Astrophys.",
    volume = "660",
    pages = "A130",
    year = "2022"
}

@article{Prantzos:2010wi,
    author = "Prantzos, N. and others",
    title = "{The 511 keV emission from positron annihilation in the Galaxy}",
    eprint = "1009.4620",
    archivePrefix = "arXiv",
    primaryClass = "astro-ph.HE",
    doi = "10.1103/RevModPhys.83.1001",
    journal = "Rev. Mod. Phys.",
    volume = "83",
    pages = "1001--1056",
    year = "2011"
}

@article{Fiorillo:2026shb,
    author = "Fiorillo, Damiano F. G. and Lucente, Giuseppe and Sakstein, Jeremy and Vitagliano, Edoardo",
    title = "{Millicharged Particle Production During Late-Stage Stellar Evolution}",
    eprint = "2604.02419",
    archivePrefix = "arXiv",
    primaryClass = "hep-ph",
    month = "4",
    year = "2026"
}

@article{Fiorillo:2026vkd,
    author = "Fiorillo, Damiano F. G. and Lucente, Giuseppe and Sakstein, Jeremy and Vitagliano, Edoardo and Cantiello, Matteo",
    title = "{The Black Hole Mass Gap as a New Probe of Millicharged Particles}",
    eprint = "2604.02413",
    archivePrefix = "arXiv",
    primaryClass = "hep-ph",
    month = "4",
    year = "2026"
}

@article{Fiorillo:2026psi,
    author = "Fiorillo, Damiano F. G. and Lucente, Giuseppe and Sakstein, Jeremy and Vitagliano, Edoardo",
    title = "{Millicharged Particle Constraints from Asymptotic Giant Branch Stars}",
    eprint = "2605.30422",
    archivePrefix = "arXiv",
    primaryClass = "hep-ph",
    month = "5",
    year = "2026"
}

@article{Holdom:1985ag,
    author = "Holdom, Bob",
    title = "{Two U(1)'s and Epsilon Charge Shifts}",
    reportNumber = "UTPT-85-30",
    doi = "10.1016/0370-2693(86)91377-8",
    journal = "Phys. Lett. B",
    volume = "166",
    pages = "196--198",
    year = "1986"
}

@article{Okun:1982xi,
    author = "Okun, L. B.",
    title = "{LIMITS OF ELECTRODYNAMICS: PARAPHOTONS?}",
    reportNumber = "ITEP-48-1982",
    journal = "Sov. Phys. JETP",
    volume = "56",
    pages = "502",
    year = "1982"
}

@article{Galison:1983pa,
    author = "Galison, Peter and Manohar, Aneesh",
    title = "{TWO Z's OR NOT TWO Z's?}",
    reportNumber = "HUTP-83/A059",
    doi = "10.1016/0370-2693(84)91161-4",
    journal = "Phys. Lett. B",
    volume = "136",
    pages = "279--283",
    year = "1984"
}

@article{Holdom:1986eq,
    author = "Holdom, Bob",
    title = "{Searching for $\epsilon$ Charges and a New U(1)}",
    reportNumber = "UTPT-86-03",
    doi = "10.1016/0370-2693(86)90470-3",
    journal = "Phys. Lett. B",
    volume = "178",
    pages = "65--70",
    year = "1986"
}

@article{Dienes:1996zr,
    author = "Dienes, Keith R. and Kolda, Christopher F. and March-Russell, John",
    title = "{Kinetic mixing and the supersymmetric gauge hierarchy}",
    eprint = "hep-ph/9610479",
    archivePrefix = "arXiv",
    reportNumber = "IASSNS-HEP-96-100",
    doi = "10.1016/S0550-3213(97)00173-9",
    journal = "Nucl. Phys. B",
    volume = "492",
    pages = "104--118",
    year = "1997"
}

@article{Goodsell:2009xc,
    author = "Goodsell, Mark and Jaeckel, Joerg and Redondo, Javier and Ringwald, Andreas",
    title = "{Naturally Light Hidden Photons in LARGE Volume String Compactifications}",
    eprint = "0909.0515",
    archivePrefix = "arXiv",
    primaryClass = "hep-ph",
    reportNumber = "IPPP-09-63, DCPT-09-126, DESY-09-123",
    doi = "10.1088/1126-6708/2009/11/027",
    journal = "JHEP",
    volume = "11",
    pages = "027",
    year = "2009"
}

@article{Albertus:2026fbe,
    author = "Albertus, Conrado and others",
    title = "{WISPedia -- the WISPs Encyclopedia}",
    eprint = "2602.09089",
    archivePrefix = "arXiv",
    primaryClass = "hep-ph",
    reportNumber = "IPPP/26/14, IFT-UAM/CSIC-26-9",
    month = "2",
    year = "2026"
}

@article{Fabbrichesi:2020wbt,
    author = "Fabbrichesi, Marco and Gabrielli, Emidio and Lanfranchi, Gaia",
    title = "{The Dark Photon}",
    eprint = "2005.01515",
    archivePrefix = "arXiv",
    primaryClass = "hep-ph",
    doi = "10.1007/978-3-030-62519-1",
    month = "5",
    year = "2020"
}

@article{Mikaelian:1978jg,
    author = "Mikaelian, Karnig O.",
    title = "{Astrophysical Implications of New Light Higgs Bosons}",
    reportNumber = "FERMILAB-PUB-78-032-T",
    doi = "10.1103/PhysRevD.18.3605",
    journal = "Phys. Rev. D",
    volume = "18",
    pages = "3605",
    year = "1978"
}

@article{Berezhiani:1999qh,
    author = "Berezhiani, Zurab and Drago, Alessandro",
    title = "{Gamma-ray bursts via emission of axion - like particles}",
    eprint = "hep-ph/9911333",
    archivePrefix = "arXiv",
    reportNumber = "DFAQ-TH-99-05, INFN-FE-12-99",
    doi = "10.1016/S0370-2693(99)01449-5",
    journal = "Phys. Lett. B",
    volume = "473",
    pages = "281--290",
    year = "2000"
}

@article{Schramm:1981mk,
    author = "Schramm, David N. and Wilson, James R.",
    title = "{Supernovae induced by axion like particles}",
    reportNumber = "EFI-81-61-CHICAGO",
    doi = "10.1086/160305",
    journal = "Astrophys. J.",
    volume = "260",
    pages = "868",
    year = "1982"
}

@article{Mori:2021pcv,
    author = "Mori, Kanji and Takiwaki, Tomoya and Kotake, Kei and Horiuchi, Shunsaku",
    title = "{Shock revival in core-collapse supernovae assisted by heavy axionlike particles}",
    eprint = "2112.03613",
    archivePrefix = "arXiv",
    primaryClass = "astro-ph.HE",
    doi = "10.1103/PhysRevD.105.063009",
    journal = "Phys. Rev. D",
    volume = "105",
    number = "6",
    pages = "063009",
    year = "2022"
}

@article{Mori:2025baz,
    author = "Mori, Kanji and Takiwaki, Tomoya and Kohri, Kazunori",
    title = "{Verifying the failing supernova constraint on dark photons with two-dimensional hydrodynamic simulations}",
    eprint = "2511.18122",
    archivePrefix = "arXiv",
    primaryClass = "astro-ph.HE",
    reportNumber = "KEK-TH-2782, KEK-Cosmo-0401",
    doi = "10.1103/d4zp-bxbn",
    journal = "Phys. Rev. D",
    volume = "113",
    number = "4",
    pages = "L041303",
    year = "2026"
}

@article{Dominguez:2025bgg,
    author = "Dom{\'\i}nguez, Inmacolata and Straniero, Oscar and Piersanti, Luciano and Giannotti, Maurizio and Mirizzi, Alessandro",
    title = "{The impact of axion-like particles on late stellar evolution - From intermediate-mass stars to core-collapse supernova progenitors}",
    eprint = "2508.17779",
    archivePrefix = "arXiv",
    primaryClass = "astro-ph.SR",
    doi = "10.1051/0004-6361/202556777",
    journal = "Astron. Astrophys.",
    volume = "702",
    pages = "A240",
    year = "2025"
}

@article{Heger:2008er,
    author = "Heger, Alexander and Friedland, Alexander and Giannotti, Maurizio and Cirigliano, Vincenzo",
    title = "{The Impact of Neutrino Magnetic Moments on the Evolution of Massive Stars}",
    eprint = "0809.4703",
    archivePrefix = "arXiv",
    primaryClass = "astro-ph",
    reportNumber = "LA-UR-07-4669",
    doi = "10.1088/0004-637X/696/1/608",
    journal = "Astrophys. J.",
    volume = "696",
    pages = "608--619",
    year = "2009"
}

@article{Cui:2026wxv,
    author = "Cui, Yanou and Li, Fengyi and Qi, Xiaolin and Shoemaker, Ian M. and Tsai, Yu-Dai",
    title = "{Direct Detection of Millicharged Particles from Supernovae}",
    eprint = "2606.11310",
    archivePrefix = "arXiv",
    primaryClass = "hep-ph",
    month = "6",
    year = "2026"
}

@article{Li:2024pcp,
    author = "Li, Changqian and Liu, Zuowei and Lu, Wenxi and Ye, Zicheng",
    title = "{Low-energy supernova constraints on millicharged particles}",
    eprint = "2408.04953",
    archivePrefix = "arXiv",
    primaryClass = "hep-ph",
    doi = "10.1007/JHEP07(2025)116",
    journal = "JHEP",
    volume = "07",
    pages = "116",
    year = "2025"
}

@article{Rrapaj:2019eam,
    author = "Rrapaj, Ermal and Sieverding, Andre and Qian, Yong-Zhong",
    title = "{Rate of dark photon emission from electron positron annihilation in massive stars}",
    eprint = "1904.10567",
    archivePrefix = "arXiv",
    primaryClass = "hep-ph",
    doi = "10.1103/PhysRevD.100.023009",
    journal = "Phys. Rev. D",
    volume = "100",
    number = "2",
    pages = "023009",
    year = "2019"
}

@article{Wang:2026zaq,
    author = "Wang, Zun and Zhang, Yongchao",
    title = "{Revisiting supernova constraints on dark gauge bosons}",
    eprint = "2610.35379",
    archivePrefix = "arXiv",
    primaryClass = "hep-ph",
    month = "10",
    year = "2026"
}

%
%


\clearpage
\appendix
\onecolumngrid
\setcounter{equation}{0}
\setcounter{figure}{0}
\setcounter{section}{0}
\setcounter{subsection}{0}
\setcounter{table}{0}
\setcounter{page}{1}
\counterwithout{equation}{section}
\renewcommand{\theequation}{S\arabic{equation}}
\renewcommand{\thefigure}{S\arabic{figure}}
\renewcommand{\thetable}{S\arabic{table}}

\begin{center}
  \textbf{\large Presupernovae Strongly Constrain Electrophilic MeV Bosons}\par
  \vspace{0.05in}
  {\itshape\large Supplemental Material}\par
  \vspace{0.05in}
  Yu Cheng, Damiano F.\ G.\ Fiorillo, Chui-Fan Kong, and Edoardo Vitagliano
\end{center}

In this supplemental material, we detail the production rates of DPs in the stellar plasma, which are calculated using the MESA simulation profiles.  
We additionally discuss the preSN spatial and mass distributions, as well as the positron survival probability.
We compare the smeared $511\,\rm keV$ photon flux against the experimental dataset, and quantify systematic uncertainties for different preSN profiles and positron smearing scales.
Finally, we extend our production rate calculations and bounds to the case of axion-like particles (ALPs).

\appendix

\section{DP production from preSN core}

\textit{Production rate.}---A DP of momentum $k$ and energy $\omega=\sqrt{k^2+ m_{A'}^2}$ mixes 
with the transverse and longitudinal 
photon modes of the plasma. The 
number of DPs produced per unit 
volume and time is~\cite{An:2013yfc,Redondo:2013lna,Hardy:2016kme,Caputo:2025avc}
\begin{equation}
    \frac{dN_{A'}}{dV\,dt}=\int\frac{d^3k}{(2\pi)^3}\,\frac{\varepsilon^2m_{A'}^4}{\omega\,(e^{\omega/T}-1)}\sum_{i=T,L} g_i\,
    \frac{|{\rm Im}\,\pi_i|}{(m_{A'}^2-{\rm Re}\,\pi_i)^2+({\rm Im}\,\pi_i)^2},
\label{eq:sm-master}
\end{equation}
where $\pi_i(\omega,k)$ are the photon self-energies of the medium, 
whose imaginary parts 
describe the production processes and whose real parts the in-medium 
mixing. Here $i=(T,L)$ denotes the 
transverse and longitudinal modes 
with $g_T=2$ and $g_L=1$, 
respectively.  
We take the expression for ${\rm Re}\,\pi_i$ from~\cite{Raffelt:1996wa}, which 
reduces to ${\rm Re}\,\pi_T\simeq\omega_{\rm pl}^2$ and ${\rm Re}\,\pi_L\simeq\omega_{\rm pl}^2 m_{A'}^2/\omega^2$ for $k\ll\omega$, 
with the plasma frequency of the relativistic and partially degenerate electron gas
\begin{equation}
    \omega_{\rm pl}^2=\frac{4\alpha}{\pi}\int_0^\infty dp\,\frac{p^2}{E}\left(1-\frac{v^2}{3}\right)\left[f_-(E)+f_+(E)\right],\qquad
    f_\mp(E)=\frac{1}{e^{(E\mp\mu_e)/T}+1},
\label{eq:sm-wp}
\end{equation}
where $E=\sqrt{p^2+m_e^2}$, $v=p/E$, 
and the electron chemical potential 
$\mu_e$ follows from charge 
neutrality at the local $\rho$ and 
$Y_e$. 
The imaginary part is the net 
damping rate of a photon-like 
excitation with the DP four-momentum, 
$|{\rm Im}\,\pi_i|=\omega(1-e^{-\omega/T})\,\Gamma_i^{\rm abs}$, with $\Gamma_i^{\rm abs}$ its absorption 
rate for unit mixing. Here the factor $1-e^{-\omega/T}$ expresses the
detailed balance with the inverse (emission) process. 
Inserting it into Eq.~\eqref{eq:sm-master}, the 
production rate away from resonances is
\begin{equation}
    \frac{dN_{A'}}{dV\,dt}=\varepsilon^2\int\frac{d^3k}{(2\pi)^3}\,e^{-\omega/T}\sum_{i=T,L} g_i\,w_i\,\Gamma_i^{\rm abs}(\omega,k),\qquad
    w_i=\left(\frac{m_{A'}^2}{m_{A'}^2-{\rm Re}\,\pi_i}\right)^2,
\label{eq:sm-channel}
\end{equation}
which is a product of
the Boltzmann factor, the 
in-medium absorption rate, and the 
mixing factor. 
Since $\omega_{\rm pl}\le0.34$~MeV 
up to the onset of infall and $\rm{Re} \pi_T\leq 3\omega_{\rm pl}^2/2$  in the Braaten–Segel approximation~\cite{Braaten:1993jw}, 
the resonance $m_{A'}^2={\rm Re}\,\pi_i$ is never met for $m_{A'}>2m_e$. 
The absorption rate is the sum of 
three processes, corresponding to 
production by coalescence $e^+e^-\to A'$, 
semi-Compton scattering $\gamma e^-\to e^-A'$, 
and bremsstrahlung $e^-Z\to e^-ZA'$.
In the following, we discuss each process.

\textit{Coalescence.}---Above the 
pair threshold, the dominant 
absorption is the decay $A'\to e^+e^-$. 
Its vacuum width is $\Gamma_{A'}$ of Eq.~\eqref{eq:GammaAp}, 
which is
the same for all polarizations and 
$\Gamma_{A'}m_{A'}/\omega$ in the plasma 
frame. Averaged over polarizations 
the decay is isotropic in the DP rest 
frame, so in the plasma frame the 
electron energy $E$ is uniform 
between $E_\pm=(\omega\pm k\beta)/2$, 
with $\beta=\sqrt{1-4m_e^2/m_{A'}^2}$, 
and the positron carries $\omega-E$. 
A DP of definite polarization decays 
with the rest-frame angular 
distribution $1+\cos^2\theta_*+(4m_e^2/m_{A'}^2)\sin^2\theta_*$ for transverse mode 
and $\sin^2\theta_*+(4m_e^2/m_{A'}^2)\cos^2\theta_*$ for longitudinal mode, 
with $E=(\omega+k\beta\cos\theta_*)/2$. The 
in-medium rate is the vacuum rate 
multiplied by the unoccupied fraction of the 
final states,
\begin{equation}
    \varepsilon^2\,\Gamma^{\rm abs}_{{\rm pair},i}=\Gamma_{A'}\,\frac{m_{A'}}{\omega}\,P_i(\omega,k),\qquad
    P_i(\omega,k)=\frac{1}{k\beta}\int_{E_-}^{E_+}dE\;W_i(E)\left[1-f_-(E)\right]\left[1-f_+(\omega-E)\right],
\label{eq:sm-pair-gamma}
\end{equation}
with the normalized angular weights
\begin{equation}
    W_T(E)=\frac{3\left[m_{A'}^2+4m_e^2+\frac{m_{A'}^2}{k^2}(2E-\omega)^2\right]}{4\left(m_{A'}^2+2m_e^2\right)},\qquad
    W_L(E)=\frac{3m_{A'}^2\left[1-\frac{(2E-\omega)^2}{k^2}\right]}{2\left(m_{A'}^2+2m_e^2\right)},\qquad
    \frac{2W_T+W_L}{3}=1,
\label{eq:sm-weights}
\end{equation}
so that $P_i\to1$ without blocking and the polarization average $P=(2P_T+P_L)/3$ is the unweighted integral. The coalescence rate is therefore
\begin{equation}
    \frac{dN_{\rm pair}}{dV\,dt}=\Gamma_{A'}\sum_{i=T,L}g_i\int\frac{d^3k}{(2\pi)^3}\,\frac{m_{A'}}{\omega}\,e^{-\omega/T}\,w_i(\omega,k)\,P_i(\omega,k)
    =\frac{\Gamma_{A'}m_{A'}}{2\pi^2}\sum_{i=T,L}g_i\int_{m_{A'}}^{\infty}d\omega\,k\,e^{-\omega/T}\,w_i\,P_i ,
\label{eq:sm-pair-rate}
\end{equation}
which with Boltzmann statistics ($P_i\to1$) and vacuum mixing ($w_i\to1$) integrates 
to $(3/2\pi^2)\Gamma_{A'}m_{A'}^2TK_1(m_{A'}/T)$, 
Eq.~\eqref{eq:emissivity_exact} of the 
main text, with mean energy $\langle\omega\rangle=m_{A'} K_2(m_{A'}/T)/K_1(m_{A'}/T)$.

\textit{Semi-Compton.}---For the semi-Compton process, $\gamma(q)+e^-(p')\to A'(k)+e^-(p)$, the same 
detailed-balance argument gives the 
production rate as $e^{-\omega/T}$ multiplied by the corresponding absorption rate ($A'(k)+e^-(p)\to\gamma(q)+e^-(p')$), summed over the DP polarizations,
\begin{equation}
    \varepsilon^2\,\Gamma^{\rm abs}_{\rm C}(\omega,k)=\frac{1}{2\omega}\int\!\frac{d^3p}{(2\pi)^3\,2E}\,f_-(E)\int d\Phi_2\,
    \sum|\mathcal{M}_{\rm C}|^2\left[1-f_-(E')\right]\left[1+f_\gamma(q)\right],
\label{eq:abs-compton}
\end{equation}
where $d\Phi_2$ is the two-body phase space of the final $\gamma(q)e^-(p')$ state and $f_\gamma=(e^{q/T}-1)^{-1}$. The amplitude has the two diagrams of Compton scattering with the DP coupled to the electron current with strength $\varepsilon e$. 
Summing over the three DP polarizations, the two photon polarizations, and both electron spins, one can obtain the matrix element in terms of $D_1=(p+k)^2-m_e^2=2p\cdot k+m_{A'}^2$ and $D_2=m_e^2-(p-q)^2=2p\cdot q$,
\begin{equation}
    \sum|\mathcal{M}_{\rm C}|^2=8e^4\varepsilon^2\left[\frac{D_2}{D_1}+\frac{D_1}{D_2}
    +2(m_{A'}^2+2m_e^2)\left(\frac{1}{D_1}-\frac{1}{D_2}+\frac{m_{A'}^2}{D_1D_2}\right)
    +2m_e^2(m_{A'}^2+2m_e^2)\left(\frac{1}{D_1}-\frac{1}{D_2}\right)^2\right].
\label{eq:sm-compton-M2}
\end{equation}
For simplicity, we neglect the 
polarization-dependent in-medium mixing in the semi-Compton contribution, setting
$w_T=w_L=1$.
This is because in the adopted Braaten–Segel approximation,
which is $0\leq {\rm Re}\pi_L\leq \omega_{\rm pl}^2$ and $\omega_{\rm pl}^2\leq {\rm Re}\pi_T \leq 3\omega_{\rm pl}^2/2$, 
a typical pre-SN plasma frequency with $\mathcal{O}(10)\,$keV gives only minor corrections to mixing $w_i$ in Eq.~\eqref{eq:sm-channel} over most of the DP mass range from $\mathcal{O}(100)\,$keV to $10\,$MeV.

\textit{Bremsstrahlung.}---For $e^-(p)+Z\to e^-(p')+Z+A'(k)$, we use the Born approximation in the static Coulomb field of an ion of charge $Z$ and density $n_Z$, screened by the electrons and the ions,
\begin{equation}
    \sum|\mathcal{M}_{\rm B}|^2=\frac{(4\pi\alpha)^3\varepsilon^2Z^2}{(|\bq|^2+k_s^2)^2}\,S(p,p',k),\qquad
    k_s^2=4\pi\alpha\,\frac{\partial(n_{e^-}-n_{e^+})}{\partial\mu_e}+\frac{4\pi\alpha\,Z^2n_Z}{T},\qquad \bq=\bp-\bp'-\bk,
\label{eq:sm-brems}
\end{equation}
where $\bq$ is the momentum 
transferred to the ion and $S$ is the 
squared amplitude summed over both 
electron spins and the three DP 
polarizations. Since the static ion 
absorbs momentum but no energy, 
the process is described by the 
invariants of the electron-DP 
subsystem, $s=(p'+k)^2-m_e^2$, $u=(p-k)^2-m_e^2$, and $t=(p-p')^2$, 
together with $|\bq|^2$, which are 
related by $s+u+t+|\bq|^2=m_{A'}^2$. 
We can then express $S$ in terms of invariants,
\begin{equation}
    S=\frac{4(s^2+u^2)}{su}
    -\frac{16|\bq|^2}{su}\left[E^2+E'^2+\frac{t+m_{A'}^2}{2}\right]
    -4(m_{A'}^2+2m_e^2)\frac{(s+u)^2}{s^2u^2}\left[4\left(\frac{uE+sE'}{s+u}\right)^2-|\bq|^2\right],
\label{eq:sm-S}
\end{equation}
with $E$, $E'$ being the electron energies in the ion rest frame.
The production rate is $n_Z\int f_-(E)[1-f_-(E')]\sum|\mathcal{M}_{\rm B}|^2\,d\Phi$ 
over the initial electron and the 
three-body final state with the 
energy of the recoilless ion fixed.

\begin{figure}[t!]
\centering
\includegraphics[height=0.35\textwidth]{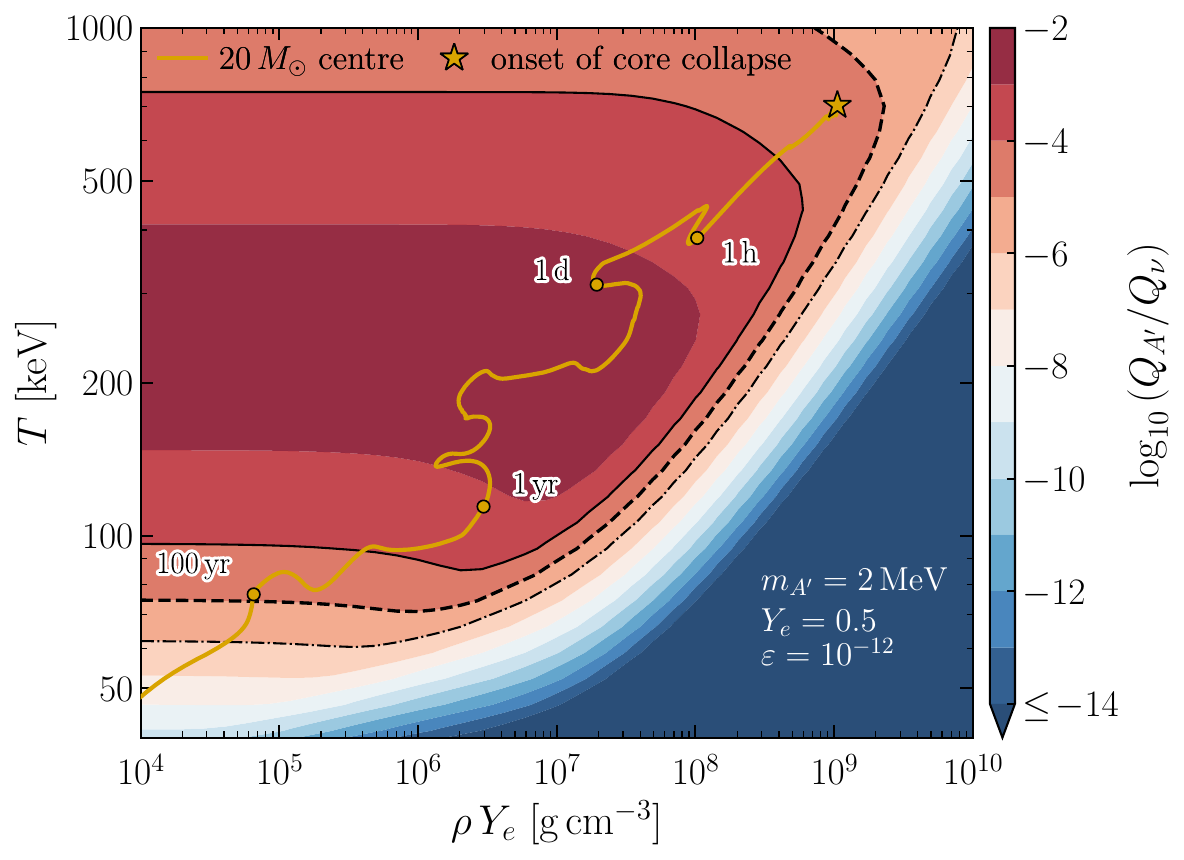}
\includegraphics[height=0.35\textwidth]{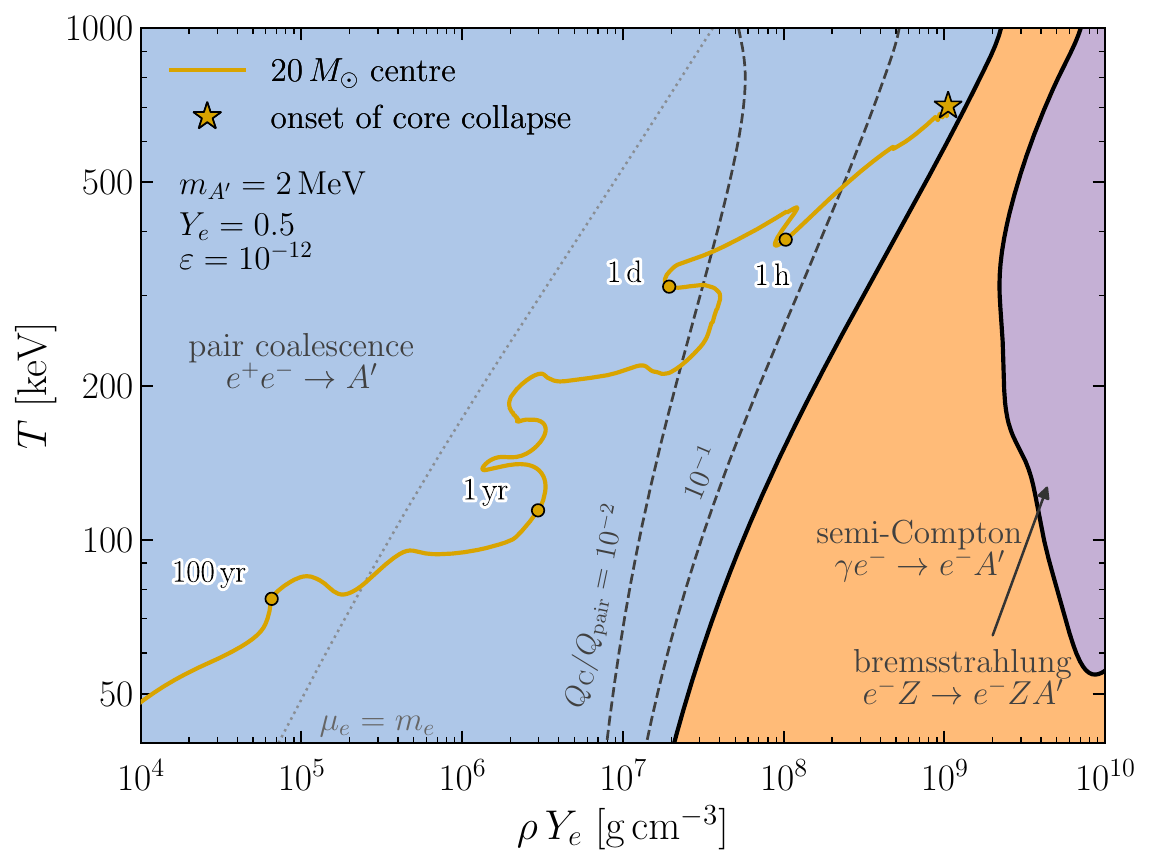}
\caption{\emph{Left}: ratio of the DP 
energy-loss rate to the neutrino energy-loss rate, $Q_{A'}/Q_\nu$, in the 
plane of electron density $\rho Y_e$ and 
temperature $T$, for $m_{A'}=2$~MeV and 
$\varepsilon=10^{-12}$. \emph{Right}: dominant DP 
production channel in the same plane for 
$m_{A'}=2\,$MeV, colored by the largest of the 
coalescence, semi-Compton, 
and bremsstrahlung emissivities. The 
dashed lines correspond to the ratio of semi-Compton to pair coalescence emissivities $Q_{\rm C}/Q_{\rm pair}=10^{-2}$ and $10^{-1}$, while the 
dotted line corresponds to the onset of degeneracy $\mu_e=m_e$.
In both panels, the yellow line is the central trajectory of the benchmark $20\,M_\odot$ star, with markers at $100$~yr, $1$~yr, $1$~d and $1$~h before core collapse and a star at the onset of infall.
}
\label{fig:sm-maps}
\end{figure}

\section{MESA simulations and stellar model comparisons}

We determine the DP spectra using radial profiles simulated from
the one-dimensional stellar evolution code Modules for
Experiments in Stellar Astrophysics (MESA) \cite{Paxton:2010ji,Paxton:2013pj}. 
We evolve stellar models from the pre-main sequence to the onset of core collapse using the \texttt{zams\_to\_cc} setup for massive stars and extract profiles at different evolutionary stages.
We simulate a grid of models between $9\,M_\odot$ and $60\,M_\odot$, with 
 $1\,M_\odot$ spacing in 
the range of $[9,30]\,M_\odot$ and 
$2\,M_\odot$ spacing in the range of $[30,60]\,M_\odot$.
The metallicity used in the simulation is taken as $Z=0.02$ for a benchmark.

In the left panel of Fig.~\ref{fig:sm-maps},
we show the ratio of the DP emissivity $Q_{A'}=\int d\omega \omega dN_{A'}/(dV dt d\omega)$ to the neutrino energy-loss rate $Q_\nu$,
taking a $20\,M_\odot$ star with $m_{A'}=2\,$MeV and $\varepsilon=10^{-12}$ for illustration.
The ratio plateaus at 
$\le 10^{-2}$ for $T\simeq150$-$300$\,keV, 
where both $Q_{A'}$ 
and the pair-neutrino 
rate carry Boltzmann 
factors, 
and drops at high 
density, where degeneracy 
suppresses the positrons.
Along the evolution trajectory of the star, 
the DP emission is much smaller compared 
with the neutrino energy-loss rate, indicating 
that the DP emission does 
not alter the stellar 
evolution significantly. 
The right panel shows the dominant DP 
production channel on the density and temperature parameter space.
The coalescence process,  which scales as $n_+n_-$ 
and is independent of density for nondegenerate 
electrons, dominates in the low density region.
The semi-Compton process, which scales as $n_e n_\gamma$, takes over at $\rho Y_e\gtrsim2\times10^{7}\,\mathrm{g/cm^3}$ for $T=40$\,keV and $\gtrsim2\times10^{9}\,\mathrm{g/cm^3}$ for $T=1\,$MeV. 
Due to the suppression by the Coulomb propagator and ion screening,
the bremsstrahlung process only dominates in a narrow band at 
$\rho Y_e\gtrsim(2$--$7)\times10^{9}\,\mathrm{g/cm^3}$.
During the whole stellar evolution, 
the coalescence process 
dominates the DP production.

\begin{figure}[!t]
\centering
\includegraphics[width=0.6\textwidth]{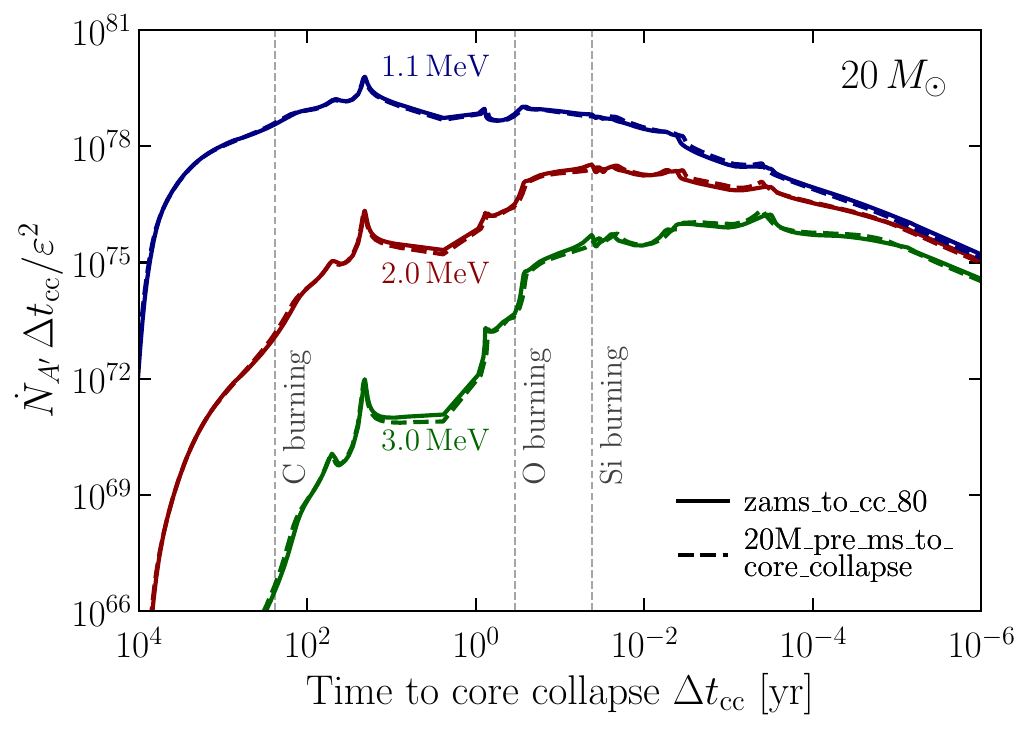}
\caption{
DP production rate multiplied by the time to core collapse and normalized by $\varepsilon^{2}$, $\dot N_{A'}\Delta t_{\rm cc}/\varepsilon^2$, for the $20\,M_\odot$, $Z=0.02$ benchmark with \texttt{zams\_to\_cc\_80} (solid) and the alternative \texttt{20M\_pre\_ms\_to\_core\_collapse} setup (dashed). The three benchmark DP masses $m_{A'}=1.1$, $2$, and $3\,$MeV are shown in different colors. 
}
\label{fig:sm-rate-compare}
\end{figure}

The integrated yield is controlled by the temperature--density track of the core during carbon, oxygen and silicon burning. To test its sensitivity to the envelope treatment, we also simulate the evolution of a $20\,M_\odot$, $Z=0.02$ star with the \texttt{20M\_pre\_ms\_to\_core\_collapse} setup.
Figure~\ref{fig:sm-rate-compare} shows $\dot N_{A'}\Delta t_{\rm cc}/\varepsilon^2$, which is the yield per unit logarithmic interval in time to collapse and divided by $\varepsilon^2$. 
The two evolution setups follow similar tracks, 
with local shifts where the burning stages ignite 
at slightly different times. 
Over $m_{A'}=1.1$--$3\,$MeV, 
the integrated yields differ by at most $6\%$,
which has a mild impact on the DP production.

\section{Spatial and mass distributions of preSNe}

To describe the probability 
distribution of Type II SNe in the 
Galaxy, it is useful to work in the 
Galactocentric coordinate system, $(r,\phi,z)$.
Here the origin is the Galactic center,
with $y$-axis directed toward the Sun, 
$r(\phi)$ the radial coordinate 
(polar angle) in the Galactic plane, 
and $z$ the height which is measured 
from the Galactic plane, and $l$ the 
Galactic longitude. For definiteness,
we take $\phi=\pi/2$ pointing toward the solar position.
The connection between
the Galactocentric coordinate system and 
the common Galactic coordinate system $(s,l,b)$ is,
\begin{subequations}
\begin{align}
    r &= \sqrt{s^2\cos^2b+d_\odot^2 - 2 d_\odot s \cos l \cos b}, \\
    z &= s\sin b.
\end{align}
\end{subequations}
where $s$ is the distance from the SN 
to the sun, $l\in [-\pi, \pi]$ 
is the Galactic longitude, $b\in [-\pi/2, \pi/2]$ is 
the Galactic latitude, and $d_\odot = 8.5\,$kpc the distance from the sun to the Galactic center.

In our study, we take the 3D  ``Ferri\`ere'' model \cite{Ferriere:2001rg} as a 
benchmark to describe the 
SN spatial distribution.
The normalized SN volume probability 
distribution can be expressed in the
Galactocentric coordinate system as,
\begin{align}
    n_{\rm F}(r,z)&\propto\left[0.79\,e^{-(z/212\,{\rm pc})^2}+0.21\,e^{-(z/636\,{\rm pc})^2}\right]
    \begin{cases}3.55\,e^{-\left[(r-3.7\,{\rm kpc})/2.1\,{\rm kpc}\right]^2}, & r<3.7\,{\rm kpc}\\[2pt]
    e^{-(r^2-d_\odot^2)/(6.8\,{\rm kpc})^2}, & r>3.7\,{\rm kpc}\end{cases}.
\label{eq:sm-ferriere}
\end{align}

We also consider an alternative 
“Lorimer” model \cite{Lorimer:2006qs} for comparison, which is parametrized by the following,
\begin{equation}
    n_{\rm L}(r,z)
    \propto
    (r/d_\odot)^{1.9}e^{-5(r-d_\odot)/d_\odot}e^{-|z|/200\,{\rm pc}}.    
\end{equation}

Note that both of the above two models describe 
an axially symmetric distribution 
of SNe centered on the Galactic Center.
In a realistic case, the axial symmetry is broken due to the existence of
spiral arms in the Galaxy. 
To take this spiral arm structure 
into account, we work in the Galactocentric coordinate system and
take the four-arm model using spiral functional forms
\cite{Steiman-Cameron:2010iuq}
\begin{equation}
    {\rm F} (r, \phi, z)
    \equiv 
    \sum_i B_i 
    e^{-|r-r_3|/\sigma_r}e^{-(\phi - \phi_i(r))^2/\delta_i^2}e^{-z^2/2\sigma_z^2},
\end{equation}
where $i$ is the index of $i$-th 
spiral arm. Here $B_i$ is the scaling 
factor which can be derived by fitting the observational data and
$\phi_i(r)=\ln(r/a_i)/\alpha_i$ is the shape function.
The values of parameters, $(B_i, a_i, \alpha_i, r_3, 
\delta_i, \sigma_z, \sigma_r)$, are given in \cite{Steiman-Cameron:2010iuq}.
Multiplying $F(r,\phi,z)$ on $n_{\rm F,L}$,
\begin{equation}
    n_{cc}(r,\phi,z)
    \equiv
    N_{\rm F,L}\times 
    n_{\rm F,L}(r,z)\times \rm{F}(r,\phi,z), 
\end{equation}
gives the SN spatial probability distribution used in our work, where $N_{\rm F,L}$ is the  corresponding normalization factor.
The above formula
satisfies the normalization condition, giving $\int n_{cc} d^3r=1$.

The mass distribution of preSN is described by the initial mass function (IMF).
In our study, we take the 
Salpeter IMF \cite{Salpeter:1955it} which
follows a power-law distribution,
\begin{equation}
    \frac{dN}{dM}
    \propto 
    M^{-\alpha},\quad \alpha = 2.35,
\end{equation}
where $dN$ is the number of 
stars formed within the 
initial mass interval $[M, M+dM]$. The Salpeter IMF
implies that low-mass stars 
are significantly more 
abundant than high-mass stars.

In our study, we adopt an initial mass range of $9 M_\odot \leq M\leq 60 M_\odot$. 
The normalized mass distribution $P(M)$ is then
\begin{equation}
    P(M)
    =
    \frac{1.35 M^{-2.35}}{M_{\rm min}^{-1.35} - M_{\rm max}^{-1.35}},
\end{equation}
where $M_{\rm min}=9\,M_\odot$ and $M_{\rm max}=60\,M_\odot$, respectively.
Here the normalization condition gives $\int_{M_{\rm min}}^{M_{\rm max}} P(M)dM =1$.

\section{Positron survival probability}

An injected positron with initial energy $E_i$ slows down
in the interstellar medium by
ionization losses and can 
annihilate with a free or bound electron 
before it stops, producing 
a $\gamma$-ray continuum instead of the 511\,keV line signal.
Therefore, in our study,
we consider the positron survival probability \cite{Dirac:1930bga,Beacom:2005qv,Cheng:2026zdc},
\begin{equation}
    P_{\rm surv}(E_i)
    =
    \exp\left[-\int_{m_e}^{E_i}dE
    \frac{n_e}{\rho}
    \frac{\sigma_{\rm ann}(E)}{S(E)}\right],
\label{eq:sm-psurv}
\end{equation}
where
\begin{equation}
    \sigma_{\rm ann}(E)
    =
    \frac{\pi\alpha^2}{m_e^2(\gamma+1)}
    \left[\frac{\gamma^2+4\gamma+1}{\gamma^2-1}
    \ln\left(\gamma+\sqrt{\gamma^2-1}\right)-\frac{\gamma+3}{\sqrt{\gamma^2-1}}\right].
\end{equation}
Here $S(E)=-\rho^{-1}dE/dx$ is the 
stopping power, 
$n_e/\rho=(X+Y/2)/m_p$ the number of electrons per unit 
mass of a neutral gas with hydrogen 
and helium mass fractions $X=0.715$ 
and $Y=0.285$, and $\sigma_{\rm ann}$ 
the Dirac cross section for $e^+e^-\to\gamma\gamma$ on an electron at rest \cite{Dirac:1930bga} 
with $\gamma=E/m_e$. 
We take stopping powers of hydrogen and helium from ESTAR \cite{NIST_ESTAR}.

The factor $\bar{P}_{\rm surv}(k)$ in Eq.~\eqref{eq:NlocProd} is the survival probability averaged over the positrons produced by DPs with fixed momentum $k$.
For $m_{A'} > 2 m_e$, the positron initial energy is $E_i = (\omega + k \beta \cos\theta_*)/2$, 
where $\omega = \sqrt{k^2 + m_{A'}^2}$, $\beta = \sqrt{1 - 4 m_e^2/m_{A'}^2}$, and $\theta_*$ is the positron emission angle in the DP rest frame relative to the boost direction. 
In the unpolarized approximation,
$E_i$ is uniformly distributed between $E_\pm = (\omega \pm k \beta)/2$, giving
\begin{equation}
    \bar{P}_{\rm surv}(k)
    = \frac{1}{2} \int_{-1}^{1} d(\cos\theta_*)\,
    P_{\rm surv}\!\left(\frac{\omega + k \beta \cos\theta_*}{2}\right)
    = \frac{1}{k \beta} \int_{E_-}^{E_+} dE_i\,P_{\rm surv}(E_i).
    \label{eq:sm-psurv-avg}
\end{equation}

\section{Smeared 511\ keV photon flux and SPI dataset}

\begin{figure}[!t]
\centering
\includegraphics[width=0.9\textwidth]{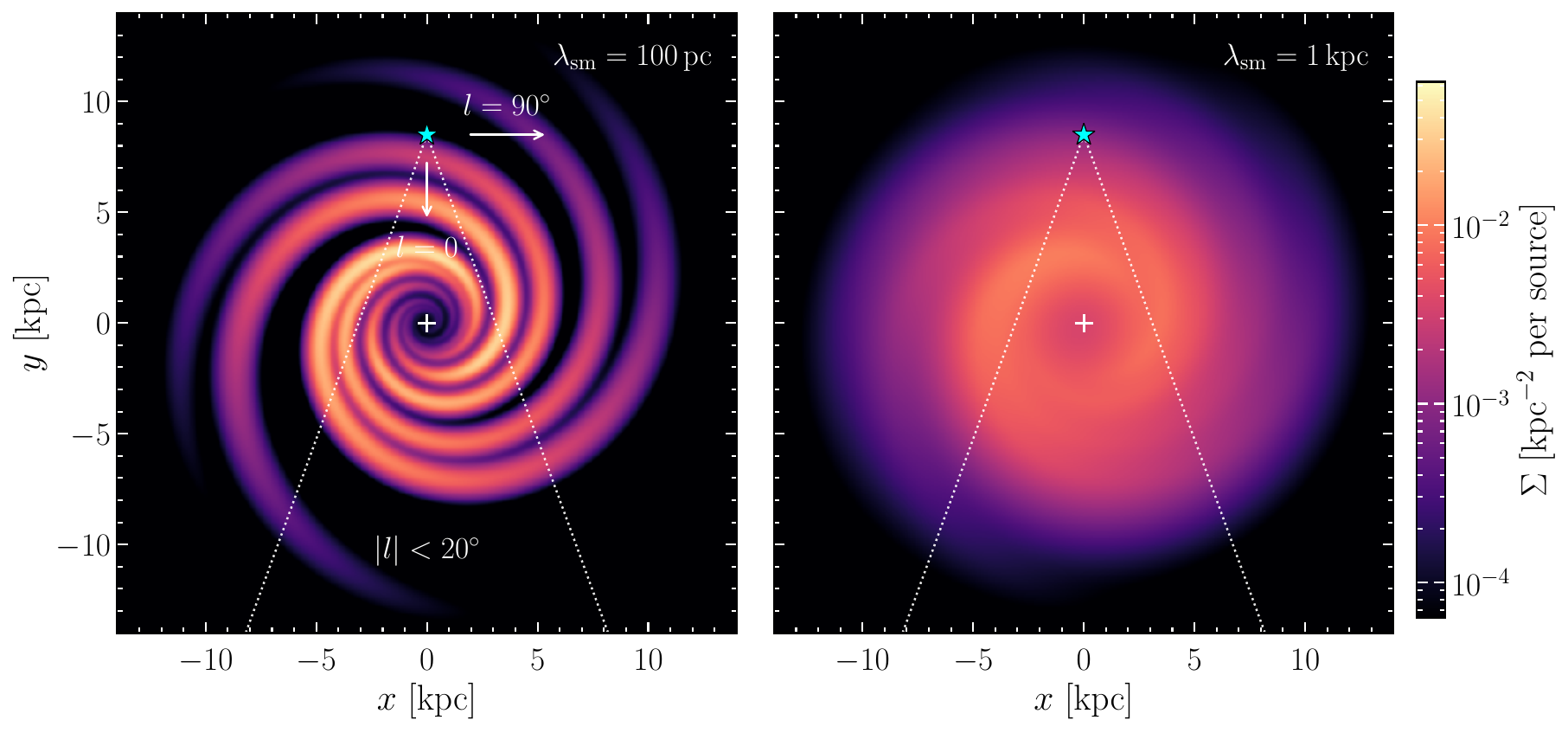}
\caption{Face-on surface density $\Sigma$ of the Ferri\`ere source distribution, normalized to one source, after smearing over $\lambda_{\rm sm}=100$~pc (left) and $1$~kpc (right). The Sun (marked by a star) is at $(x,y)=(0, 8.5)\,$kpc
and the Galactic center is marked by the cross. The dotted lines correspond to the range of $|l|<20^\circ$.
}
\label{fig:sm-smearing}
\end{figure}

The 511\ keV intensity is given by
Eq.~\eqref{eq:flux} in the 
main text. 
Following \cite{Calore:2021klc}, we introduce a 
scale parameter $\lambda_{\rm sm}$ on 
$r$ and $z$
to describe the smearing effect.
Consequently, the SN spatial probability distribution will be smeared from its source distribution $n_{cc}$ to
\begin{equation}
    \tilde{n}_{cc}
    =
    \frac{1}{4\pi \lambda_{\rm sm}^3}
    \int r'dr'\int d\phi' \int dz' n_{cc}(r',\phi',z')
    \exp\left[-\frac{\varrho}{\lambda_{\rm sm}}\right]\exp\left[-\frac{|z-z'|}{\lambda_{\rm sm}}\right],
\end{equation}
where two smearing functions are applied in the 
Galactic plane and along the vertical direction, respectively. 
Here $\varrho = \sqrt{r^2+r'^2-2rr'\cos(\phi-\phi')}$ denotes the
distance between two points on the Galactic plane.


Fig.~\ref{fig:sm-smearing} 
illustrates how DP decay length and positron propagation 
reshape the spatial distribution of 
annihilation sites.
In this figure, we define  
the face-on surface density of the 
smeared Galactic source distribution,
$\Sigma(r,\phi)\equiv \int dz \tilde{n}_{cc}(r,\phi,z) $, which is
obtained by integrating the 
smeared spatial probability distribution over the vertical 
coordinate $z$ and normalized to 
one source.
For $\lambda_{\rm sm}=100\,\mathrm{pc}$, the spiral 
structure remains prominent, and 
annihilations closely trace the 
progenitor population. Increasing the 
smearing scale to $1\,\mathrm{kpc}$ 
spreads positrons into the inter-arm 
regions, broadening the arm features 
and reducing their contrast.

\begin{figure}[!t]
\centering
\includegraphics[width=0.72\textwidth]{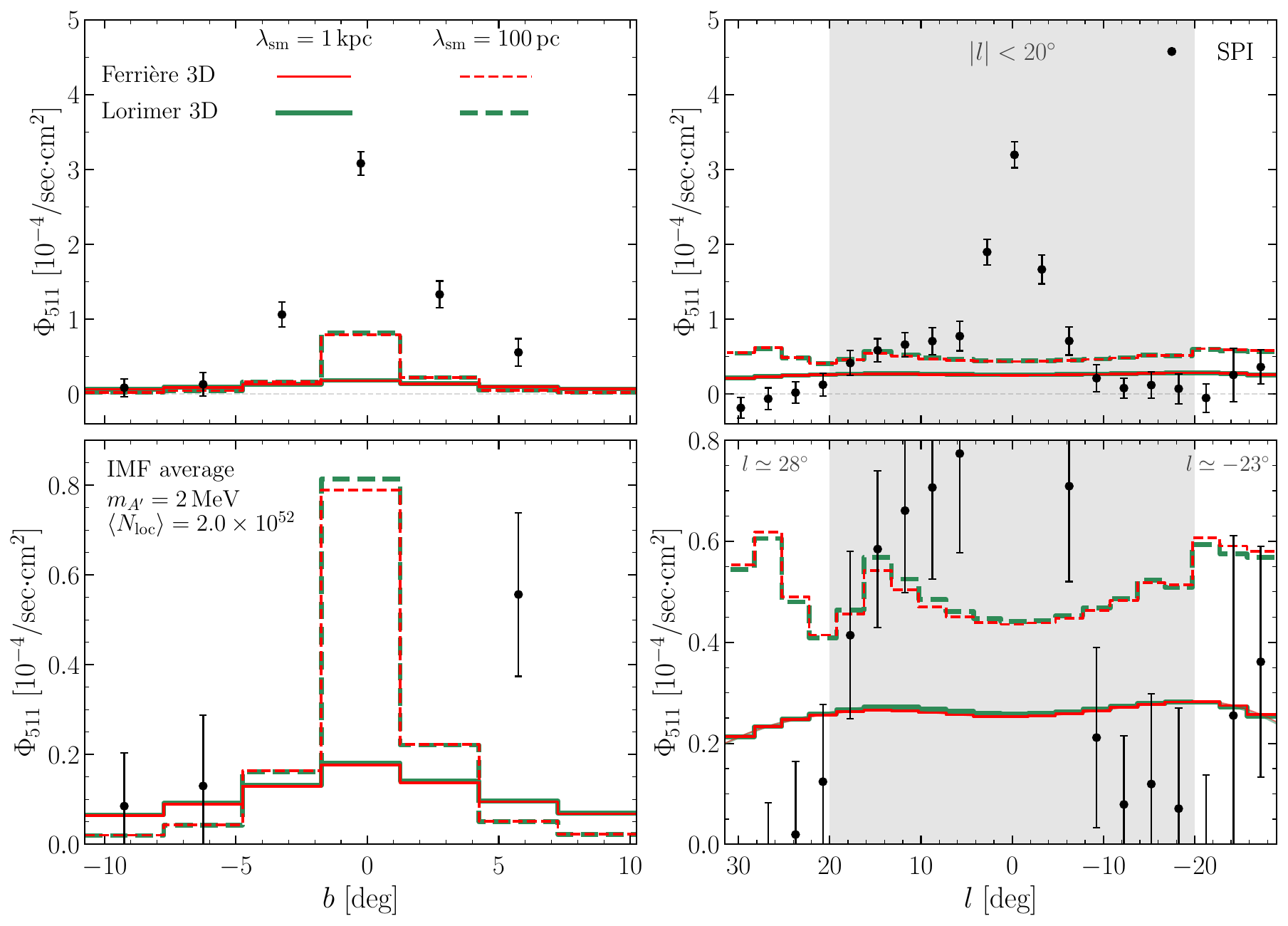}
\caption{SPI latitude (left) and longitude (right) profiles of the 511\ keV line~\cite{Siegert:2015knp} compared with the IMF-averaged preSN prediction for $m_{A'}=2$~MeV and $\langle N_{\rm loc}\rangle =2.0\times10^{52}$. 
The data are shown by points with $1\sigma$ error bars.
We take the Ferri\`ere 3D (red) and Lorimer 3D (green) source models with $\lambda_{\rm sm}=1$~kpc (solid) and $100$~pc (dashed). The lower panels zoom in at the disk level. The shaded regions in the right panels correspond to $|l|<20^\circ$.
}
\label{fig:sm-binbybin}
\end{figure}

By integrating the smeared 511\ keV intensity over each Galactic 
latitude and longitude bin, we can obtain 
the corresponding 511\ keV photon flux.
In Fig.~\ref{fig:sm-binbybin},
we show the integrated 511\,keV photon flux in the latitude binning (left) and longitude binning (right),
together with the data observed by the spectrometer on INTEGRAL (SPI) \cite{Siegert:2015knp}.
Here we take $m_{A'}=2\,$MeV and $\langle N_{\rm loc} \rangle = 2\times 10^{52}$ as a benchmark.

The predicted signal follows the 
Galactic disk, producing a pronounced 
concentration toward low latitudes 
and a relatively flat longitude profile within $|l|<20^\circ$.
In contrast,
the observed emission peaks strongly toward the Galactic center and decreases on either side.
Consequently, the one-sided criterion in Eq.~\eqref{eq:criterion} is 
most sensitive to the longitude bins 
with low measured flux away from the 
central peak, particularly those at 
negative longitudes within the 
selected region.
Increasing the smearing length $\lambda_{\rm sm}$ slightly
broadens the latitude distribution 
and reduces the flux in the inner 
bins at fixed positron yield, 
allowing a larger $\langle N_{\rm loc} \rangle$ and weakening the constraint.

For comparison, we also show the two spatial 
distribution modes, Ferri\`ere 3D and 
Lorimer 3D, in red and 
green colors, respectively. The 
predicted 511\ keV flux is weakly sensitive to the choice of spatial distribution 
models, indicating that the uncertainty of spatial distribution does not lead to a large difference in the constraint results, as we will show below.

\section{Comparison of DP constraints from different preSNe profiles and positron smearing scales}

\begin{figure}[!t]
\centering
\includegraphics[width=0.6\textwidth]{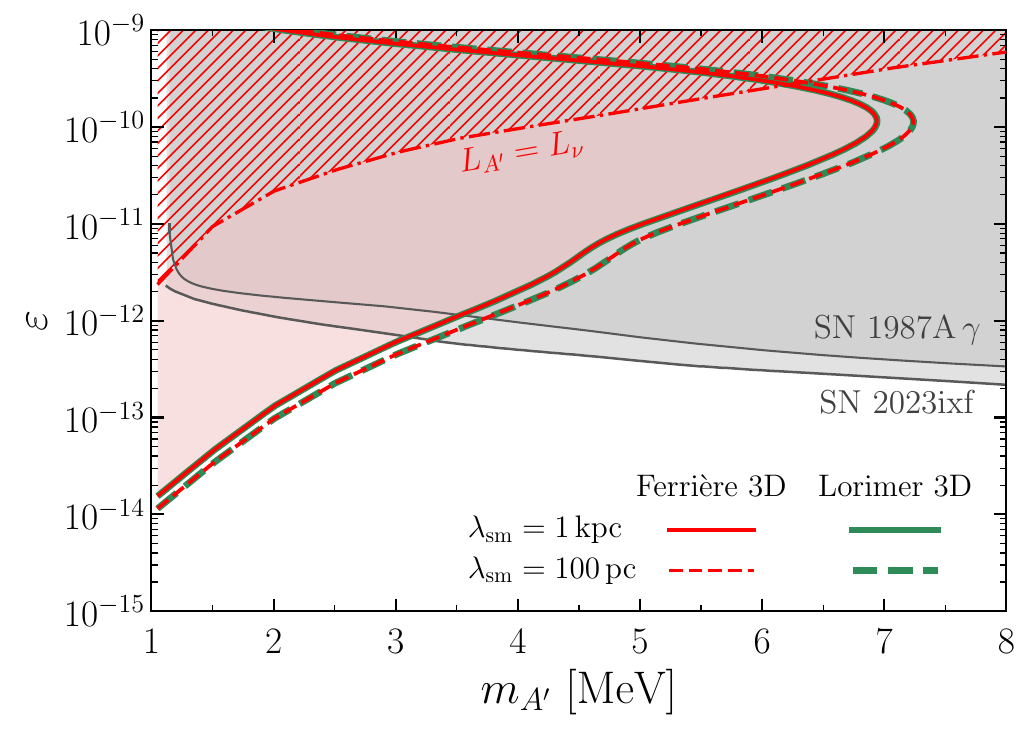}
\caption{Excluded region for the Ferri\`ere 3D model with $\lambda_{\rm sm}=1$~kpc (red solid) and $100$~pc (red dashed), 
and for the Lorimer 3D model with the same two 
choices (green). The dash-dotted line is $L_{A'}=L_\nu$, 
above which the band is hatched, and the gray regions are the SN~1987A $\gamma$-ray~\cite{Caputo:2025avc} and SN~2023ixf~\cite{Cheng:2026ugs} limits.
}
\label{fig:sm-bounds-models}
\end{figure}

In this section, we show the comparison of constraint results from different preSNe profiles and the smearing scales.
Fig.~\ref{fig:sm-bounds-models} 
compares the DP exclusion regions 
obtained with the Ferrière (red) and 
Lorimer (green) Galactic source distributions 
and two positron smearing scales, $\lambda_{\rm sm}=1\,$kpc (solid) and $\lambda_{\rm sm}=100\,$pc (dashed).
The constraints from the two spatial 
distribution models are nearly 
indistinguishable, in agreement with the predicted 511\ keV photon flux as shown above.
In contrast, the smearing effect that comes from DP decay length and 
positron propagation leads to a more 
noticeable effect. 

\section{ALP production and constraints}

For an ALP $a$ with the pseudoscalar coupling,
$\mathcal{L}\supset (g_{ae}/2m_e)
\partial_\mu a
\bar e\gamma^\mu\gamma_5e$,
the decay width for $a\to e^+e^-$ is $\Gamma_a=g_{ae}^2m_a\beta/8\pi$,
where $\beta=\sqrt{1-4m_e^2/m_a^2}$.
Using the detailed-balance argument in
the section ``DP production from preSN core'', we can derive the
coalescence rate in the axion case,
\begin{equation}
    \frac{dN_a}{dVdt}
    =
    \Gamma_a\int\frac{d^3k}{(2\pi)^3}
    \frac{m_a}{\omega}e^{-\omega/T}P(\omega,k)
    =
    \frac{g_{ae}^2m_a^2}{16\pi^3}\int_{m_a}^{\infty}d\omega\int_{E_-}^{E_+}dE f_-(E) f_+(\omega-E)
\end{equation}
where $P=(2P_T+P_L)/3$ is 
the unweighted Pauli factor of 
Eq.~\eqref{eq:sm-pair-gamma}.
The semi-Compton process, $\gamma e^-\to e^- a$, and the bremsstrahlung
process, $e^-Z\to e^-Za$, are 
evaluated with the amplitudes given in \cite{Fiorillo:2025sln}.

\begin{figure}[!t]
\centering
\includegraphics[height=0.35\textwidth]{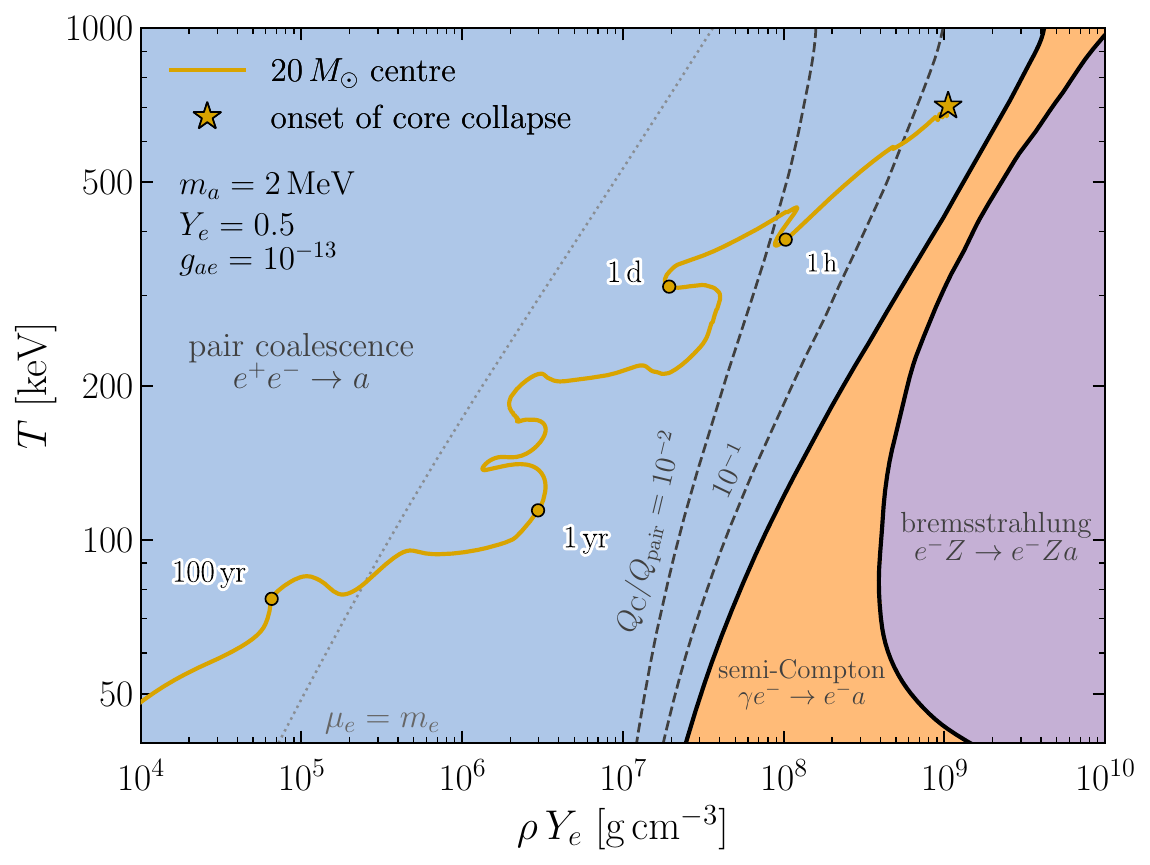}
\includegraphics[height=0.35\textwidth]{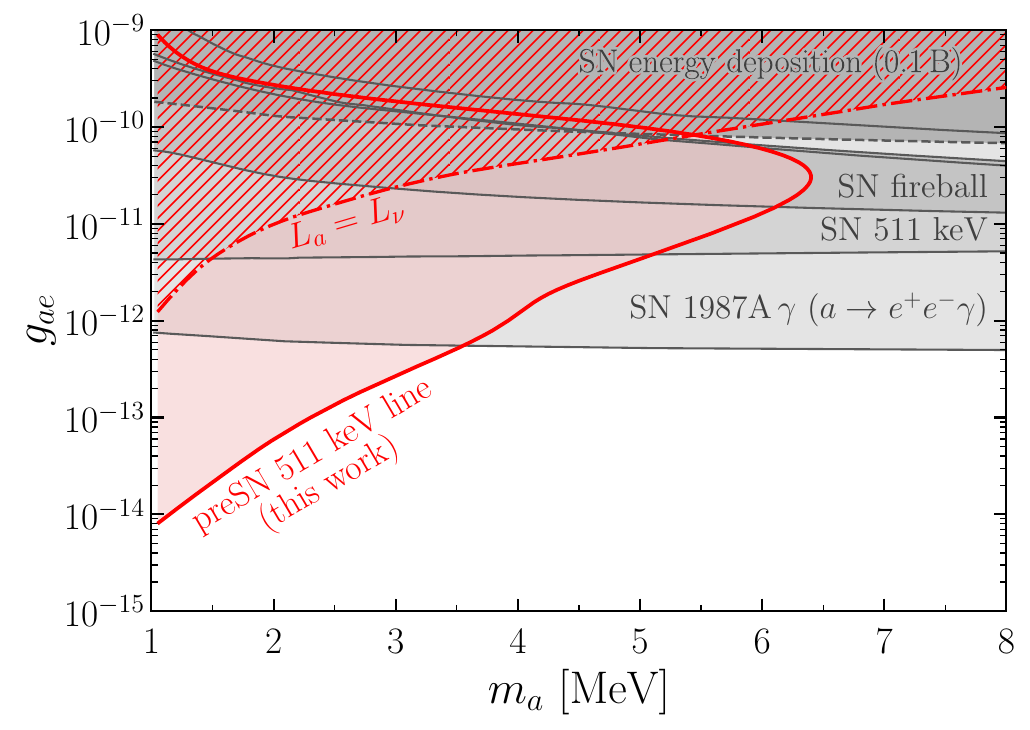}
\caption{\emph{Left:} same as the right panel of Fig.~\ref{fig:sm-maps} but for the ALP production with coupling $g_{ae}=10^{-13}$. 
\emph{Right:} region of the $(m_a,g_{ae})$ plane excluded by the Galactic 511-keV line from ALPs emitted by presupernovae. We take the Ferri\`ere 3D source model with $\lambda_{\rm sm}=1$~kpc for illustration. 
The dash-dotted line is $L_a=L_\nu$, above which the region is hatched. The gray regions are the SN bounds of Ref.~\cite{Fiorillo:2025sln} (cold SN model).
}
\label{fig:sm-alp-bounds}
\end{figure}

The left panel of Fig.~\ref{fig:sm-alp-bounds} shows the dominant ALP production channel on the 
density and temperature
parameter space.
Similar to the DP case,
in the low density region,
the coalescence process dominates.
As the density increases beyond $\rho Y_e \simeq 3\times 10^7\,{\rm g/cm}^3$ at $T=50\,$keV and $4\times 10^9\,{\rm g/cm}^3$ at $T=1\,$MeV,
the semi-Compton process starts dominating.
The bremsstrahlung dominates beyond $\rho Y_e \simeq (4{\rm -}9)\times 10^8\,{\rm g/cm}^3$ 
for $T\lesssim 300\,$keV and a few $10^9\,{\rm g/cm}^3$ at higher
temperatures.
Still, the coalescence process dominates during the
stellar evolution. 
In the right panel, 
we show the constraints on ALP parameter space.
Using the 511\,keV line, 
our result extends the existing 
constraints most substantially at low masses and small couplings.


\end{document}